\documentclass[11pt]{article}

\usepackage[preprint]{acl}

\usepackage{times}
\usepackage{latexsym}
\usepackage{booktabs}
\usepackage{multirow}
\usepackage{makecell}
\usepackage{xcolor}
\definecolor{PaperRed}{RGB}{150,40,40}
\definecolor{PaperGreen}{RGB}{30,120,80}
\definecolor{PaperGray}{gray}{0.35}
\usepackage[table]{xcolor} 
\usepackage[T1]{fontenc}
\usepackage[utf8]{inputenc}

\usepackage{microtype}
\usepackage{amsmath} 
\usepackage{graphicx}
\usepackage{subcaption}
\usepackage{inconsolata}

\usepackage{graphicx}

\usepackage[most]{tcolorbox}
\definecolor{uclablue}{RGB}{39, 116, 174}
\newtcolorbox{findings}[1][]{
	float,
  	title=#1,
        colframe=uclablue,
        top=1pt,           % 控制顶部空白
        bottom=1pt,        % 控制底部空白
        left=0pt,          % 控制左边空白
        right=0pt,          % 控制右边空白
        before skip=0.65em, after skip=0.75em,
}

\title{When KV Cache Reuse Fails in Multi-Agent Systems:\\ Cross-Candidate Interaction is Crucial for LLM Judges}

\author{
 \textbf{Sichu Liang \textsuperscript{1}}\thanks{Equal Contribution},
 \textbf{Zhenglin Wang \textsuperscript{1}}\footnotemark[1],
 \textbf{Jiajia Chu\textsuperscript{2}},
 \textbf{Pengfei Xia\textsuperscript{2}},
 \textbf{Hui Zang\textsuperscript{2}}\thanks{Corresponding author},
 \textbf{Deyu Zhou \textsuperscript{1}}\footnotemark[2]
\\
 \textsuperscript{1}Southeast University\quad
\textsuperscript{2}Huawei Technologies Ltd
}

\begin{document}
\maketitle
\begin{abstract}
Multi-agent LLM systems routinely generate multiple candidate responses that are aggregated by an LLM judge.
To reduce the dominant prefill cost in such pipelines, recent work advocates KV cache reuse across partially shared contexts and reports substantial speedups for generation agents.
In this work, we show that these efficiency gains do not transfer uniformly to \emph{judge-centric} inference.
% We study settings where a judge must \emph{jointly} compare multiple candidates within a single context and select the best one, under two candidate-generation regimes and different candidate orderings.
Across GSM8K, MMLU, and HumanEval, we find that reuse strategies that are effective for execution agents can severely perturb judge behavior: \emph{end-task accuracy may appear stable, yet the judge's selection becomes highly inconsistent with dense prefill.}
We quantify this risk using Judge Consistency Rate (JCR) and provide diagnostics showing that reuse systematically weakens cross-candidate attention, especially for later candidate blocks.
Our ablation further demonstrates that explicit cross-candidate interaction is crucial for preserving dense-prefill decisions.
Overall, our results identify a previously overlooked failure mode of KV cache reuse and highlight judge-centric inference as a distinct regime that demands dedicated, risk-aware system design.
\footnote{The code is available in \url{https://anonymous.4open.science/r/kv_reuse_fails-5B0C}}
\end{abstract}
\section{Introduction}

\begin{figure}[t]
    \centering
    \includegraphics[width=0.88\linewidth]{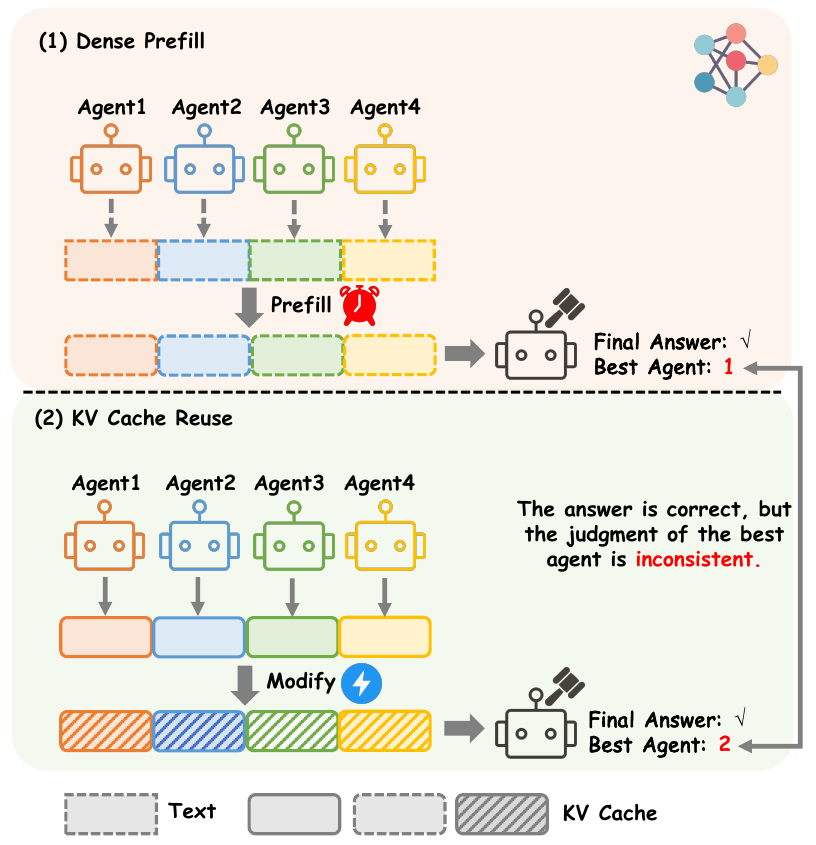}
    \caption{Illustration of decision non-invariance under judge-side KV cache reuse. \textbf{Top}: dense prefill recomputes the judge KV cache and selects Agent~1. \textbf{Bottom}: KV reuse stitches/modifies cached KV blocks, keeping the final answer correct but changing the selected best agent, despite identical candidate texts.}
    \label{fig:intro}
\end{figure}

Modern intelligent systems often reason by considering multiple alternatives rather than committing to a single trajectory---a pattern observed in human decision-making and in AI systems that rely on search, sampling, and planning~\cite{churchland2008decision,browne2012survey,silver2016mastering}. 
\textbf{Large language models (LLMs)} have begun to exhibit similar behavior through techniques such as multi-sample decoding~\cite{leviathan2023fast,gui2024bonbon,sun2024fast}, self-consistency~\cite{wang2023selfconsistency,li-etal-2025-revisiting-self}, preference modeling~\cite{rafailov2023direct,ouyang2022training,sun2025rethinking,ye2025learning}, and search-based inference~\cite{yao2023tree,besta2024graph}, where multiple candidate solutions are generated and subsequently evaluated.

LLM-based \textbf{multi-agent systems (MAS)} \cite{hong2024metagpt, wu2024autogen} represent a concrete instantiation of this paradigm, where multiple execution agents generate diverse candidate outputs that are aggregated by a central meta-agent~\cite{pmlr-v267-zhang25bc} responsible for planning, coordination or \textbf{judging}~\cite{chen2024agentverse, fourney2024magentic, zhang2025agentorchestra}. 
To enable such collaboration, execution agents and the meta-agent must repeatedly exchange intermediate reasoning, candidate outputs, and contextual information, leading to substantial overlap in their effective input contexts\cite{wang-etal-2025-agentdropout,zhang2025cut}. 
While this interaction improves robustness and reasoning quality, it also introduces severe computational redundancy: each agent repeatedly performs a full prefill pass over largely overlapping context, causing inference cost to scale rapidly with the number of agents~\cite{KVCOMM}.

A natural solution is \textbf{KV cache reuse}~\cite{vllm, zheng2024sglang}. 
During LLM inference, the prefill stage encodes the entire input context and constructs key--value (KV) caches for attention, which can be reused directly when multiple inputs share overlapping prefixes. 
However, in multi-agent settings, simple prefix-based reuse is often ineffective: although agents typically share substantial contextual overlap (e.g., task instructions or exchanged messages), their prompts frequently diverge due to role specifications, intermediate reasoning, or agent-specific context. 
Consequently, KV caches across agents are similar but not identical, preventing direct reuse~\cite{zhao2025smallkv,liu2025speculative}. 
This has motivated \emph{cross-prefix} (approximate) reuse methods that enable reuse under partially shared prefixes and have reported substantial speedups for generation agents in multi-agent pipelines~\cite{kvlink,CacheBlend,KVCOMM,turborag}.

Most existing studies evaluate KV reuse primarily from the perspective of execution agents, implicitly assuming that efficiency gains transfer uniformly across all components of a multi-agent pipeline. 
Yet the impact of KV reuse on \emph{judge} agents---which must compare multiple candidates jointly and produce a selection~\cite{li2024llms}---remains largely unexplored.

This gap is non-trivial, as judging constitutes a qualitatively different inference regime from generation~\cite{li-etal-2025-generation}. 
Prior work has examined the reliability of LLM-based judges~\cite{zheng2024large}, documenting systematic biases such as position preference~\cite{li-etal-2024-split} and sensitivity to irrelevant contexts~\cite{shi2023large}. 
However, these failures are typically attributed to model limitations or task-level biases, rather than to \emph{system-level efficiency optimizations} that modify the inference procedure. 
Consequently, how KV cache reuse alters judge behavior in multi-candidate comparison settings remains an open question.
As shown in Figure~\ref{fig:intro}, KV reuse may keep the final answer unchanged while altering which candidate the judge selects, motivating judge-centric diagnostics beyond accuracy.
% However, existing studies predominantly evaluate KV reuse from the perspective of execution agents, implicitly assuming that efficiency gains transfer uniformly across all components of the system. 
% In particular, the impact of KV reuse on judge agents—responsible for cross-candidate comparison and attribution~\cite{li2024llms}—remains largely unexplored.

% This gap is non-trivial, as judging constitutes a qualitatively different inference regime from generation\cite{li-etal-2025-generation}. 
% Notably, recent studies have examined the reliability of LLM-based judges themselves~\cite{zheng2024large}, revealing systematic biases such as position preference~\cite{li-etal-2024-split} and sensitivity to generation strategies in candidate selection~\cite{shi2023large}. 
% However, these works primarily attribute judge failures to model reasoning limitations or surface-level biases inherent to the evaluation task, rather than to efficiency optimizations applied at the system level. 
% As a result, how KV cache reuse alters judge behavior in multi-candidate comparison settings remains an open question.

In this paper, we revisit KV cache reuse from a \emph{judge-centric} perspective, where an LLM judge must \emph{jointly} compare multiple candidates within a single context and output both a final answer and the selected candidate.
We show that judge-side reuse can be \emph{decision-non-invariant}: it may preserve end-task accuracy while substantially changing which candidate the judge selects, especially under order perturbations.
These results highlight judge-centric inference as a distinct regime in which preserving cross-candidate interactions is critical.

Our key contributions are:
\begin{itemize}
    \item \textbf{Judge-centric evaluation of KV reuse.} We provide a systematic study of KV reuse in multi-candidate judging across tasks, candidate-generation regimes, and candidate orderings.
    % with execution-side generation held fixed to isolate judge-side effects.
    \item \textbf{Decision instability beyond accuracy.} We introduce Judge Consistency Rate (JCR) to quantify decision non-invariance relative to dense prefill, and show that Acc. and JCR can be decoupled---accuracy may appear stable while selection behavior becomes inconsistent under reuse.
    \item \textbf{Mechanistic diagnostics and implications.} Through attention analyses and controlled ablations, we attribute the failure to disrupted cross-candidate interaction; we further develop PAL-KV as a probe to rule out agent identity as the dominant bottleneck and discuss practical directions such as interaction-aware reuse and risk-aware gating.
\end{itemize}

\section{Related Work}
\paragraph{KV Cache Reuse. }
KV cache reuse is a widely studied approach for reducing the high prefill cost of large language model (LLM) inference. Early systems~\cite{vllm,zheng2024sglang} focus on prefix caching, where KV tensors are reused only when the input context remains identical.
Recent work extends KV reuse along two main directions. One line studies \emph{cross-model} KV sharing, enabling cache reuse across model variants derived from the same backbone~\cite{droidspeak,mobilora}.
Most closely related to our study is the second line on \emph{cross-prefix} KV reuse, which enables approximate reuse when prompts share partial prefixes but are not identical. 
Methods such as \emph{TurboRAG}~\cite{turborag}, \emph{KVLink}~\cite{kvlink}, and \emph{CacheBlend}~\cite{CacheBlend} achieve this via position adjustment, linking tokens, or selective recomputation, and are typically evaluated in settings with limited cross-segment interactions. 
\emph{KVCOMM}~\cite{KVCOMM} further brings cross-prefix reuse to \emph{multi-agent systems} by using anchor examples to correct cache deviations across agent-specific prompts.

We study KV reuse in \emph{judge-centric} multi-candidate inference and show that reuse strategies effective for agent-side generation can be decision-non-invariant for judges, revealing a failure mode overlooked by prior work.

% In contrast, we study KV reuse in \emph{judge-centric} multi-candidate inference, where a meta-agent must jointly compare multiple candidates within a single context. 
% We show that reuse strategies effective for agent-side generation can be decision-non-invariant for judges, revealing a failure mode overlooked by prior work.

\paragraph{LLM-Based Multi-Agent Systems. }
% LLM-based multi-agent systems solve complex tasks by coordinating multiple specialized agents that interact through structured communication graphs~\cite{chen2024agentverse, zhuge2024gptswarm}. Prior work explores diverse agent topologies, where the \emph{execution agents} generate candidate outputs under different roles or prompts~\cite{wang-etal-2025-agentdropout, zhang2025cut}.
% Despite differences in graph design, a common and practically important pattern is the presence of a central \emph{meta-agent} that aggregates, compares, or verifies multiple execution agents' outputs to produce a final decision~\cite{li2023camel, hong2024metagpt, pmlr-v267-zhang25bc}. This meta-agent is often implemented as an LLM itself and must attend over multiple agent responses within a single context window, making it a critical component for both quality and efficiency.
LLM-based multi-agent systems coordinate multiple specialized agents through structured communication graphs~\cite{chen2024agentverse,zhuge2024gptswarm}, where \emph{execution agents} produce candidate outputs under different roles or prompts~\cite{wang-etal-2025-agentdropout,zhang2025cut}. 
A common pattern is a central LLM \emph{meta-agent} that aggregates and verifies multiple agent responses within a single context window to produce the final decision~\cite{li2023camel,hong2024metagpt,pmlr-v267-zhang25bc}, making it a key component for both quality and efficiency.

We show that KV reuse, while effective for execution agents, is not always \emph{behavior-preserving} for judge agents, often changing which candidate is selected relative to dense prefill.

\paragraph{LLM-as-a-Judge. }
% LLM-based judges are increasingly deployed as \emph{decision modules} in high-stakes downstream workflows including healthcare, finance, and legal analysis~\cite{xie-etal-2024-doclens,xie2023pixiu,raju-etal-2024-constructing}, where multiple candidate outputs must be audited, ranked, or selected before execution.
% In these applications, the judge’s selection behavior directly affects both the final outcome and the system’s accountability, making robustness and faithfulness of the judging step a first-order concern.
LLM-based judges are increasingly deployed as \emph{decision modules} in high-stakes workflows including healthcare, finance, and legal analysis~\cite{xie-etal-2024-doclens,xie2023pixiu,raju-etal-2024-constructing}, where selection robustness is critical.
Most existing formulations treat judging as either \emph{pairwise comparison} or \emph{scalar scoring}, where candidates are evaluated independently or in pairs \cite{sun2024fast, sun2025rethinking}. In contrast, many practical applications require a judge to process \emph{multiple candidates jointly} within a single context and directly select the best output \cite{tran-etal-2025-rare, gera-etal-2025-justrank}. This setting is especially common in multi-agent systems, where a judge must aggregate and attribute outputs from several execution agents.

Such judges require cross-candidate comparison and relative reasoning, unlike standard generation~\cite{li-etal-2025-generation}. We study KV cache reuse under this joint multi-candidate judging setting and show it can change the judge's selection relative to dense prefill.

% \section{Methodology}
\section{Preliminaries}

\begin{figure*}[t]
    \centering
    \includegraphics[width=0.88\linewidth]{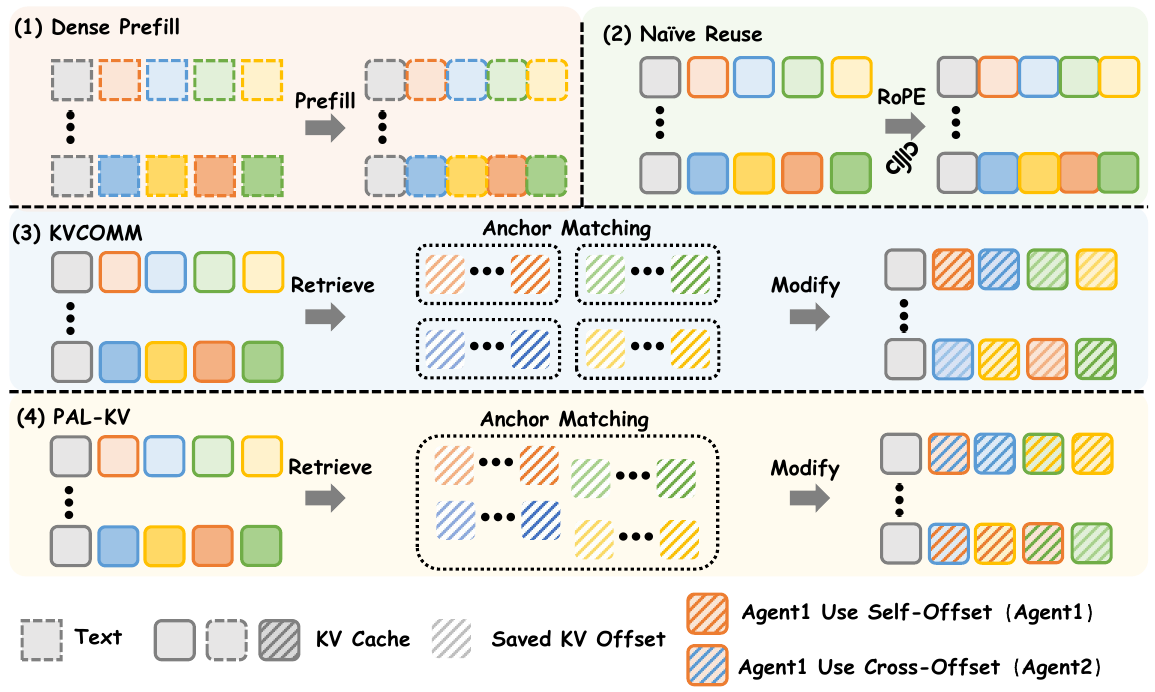}
    \caption{Judge-side KV cache construction for multi-candidate judging. \textbf{Dense prefill} recomputes the full judge cache, while \textbf{Na\"ive Reuse} aligns and stitches execution-side candidate KV chunks. \textbf{KVCOMM} retrieves anchor-based cache offsets to correct reused chunks, and \textbf{PAL-KV} pools anchors across agents for offset retrieval.}
    \label{fig:method}
\end{figure*}

% \subsection{Preliminaries}
% \label{sec:task}
% We study \emph{judge-centric} multi-candidate inference with $N$ execution agents and a central judge agent.
% Given an input question $x$, each execution agent $A_i$ uses an agent-specific prompt $p_i$ and produces a candidate response $y_i$.
% The judge agent $J$ with prompt $p_J$ then takes the set of candidates as input and outputs (i) a final answer $\hat{y}$ and (ii) the index of the selected candidate $\hat{i}$:
% \begin{equation}
% (\hat{y}, \hat{i}) \;=\; J\!\left(p_J,\;x,\; y_{\pi(1:N)}\right), \qquad \hat{i}\in\{1,\dots,N\},
% \end{equation}
% where $\pi$ denotes the presentation order of candidates in the judge input.
% Unlike pairwise comparison or scalar scoring, the judge must \emph{jointly} reason over multiple candidates within a single context, performing explicit cross-candidate comparison and selection.

\label{sec:task}
We study \emph{judge-centric} multi-candidate inference with $N$ execution agents and a central judge agent.
Given an input question $x$, each execution agent $A_i$ is instantiated as a large language model with an agent-specific prompt $p_i$ (e.g., role description or instruction), and produces a candidate response $y_i$.
A judge agent $J$ with a judge-specific prompt $p_J$ then takes all candidate responses as input and outputs (i) a final answer $\hat{y}$ and (ii) the index of the selected candidate $\hat{i}$:
\begin{equation}
(\hat{y}, \hat{i}) \;=\; J\!\left(p_J,\;x,\; y_{1:N}\right), \qquad \hat{i}\in\{1,\dots,N\}.
\end{equation}
Unlike pairwise comparison or scalar scoring, the judge must \emph{jointly} reason over multiple candidates within a single context, performing explicit cross-candidate comparison and selection.

\paragraph{Candidate generation regimes.}
We consider two common ways to construct multi-candidate inputs in multi-agent pipelines.

\textbf{(1) Progressive refinement.}
Candidates are generated sequentially, where later agents may condition on earlier candidates:
\begin{equation}
y_i = A_i\!\left(p_i,\;x,\;y_{1:i-1}\right), \qquad i=1,\dots,N.
\label{eq:progressive}
\end{equation}

\textbf{(2) Parallel exploration.}
Candidates are generated independently without observing each other:
\begin{equation}
y_i = A_i\!\left(p_i,\;x\right), \qquad i=1,\dots,N.
\label{eq:parallel}
\end{equation}

\paragraph{Judge input ordering.}
To disentangle content-based judging from potential ordering effects, we optionally randomize the presentation order of candidates to the judge.
Let $\pi$ be a permutation over $\{1,\dots,N\}$.
The judge observes candidates in the permuted order $y_{\pi(1:N)}$:
\begin{equation}
(\hat{y},\hat{i}) = J\!\left(p_J ,\;x,\; y_{\pi(1:N)}\right),
\label{eq:permute}
\end{equation}
where \texttt{no-shuffle} uses the identity permutation and \texttt{shuffle} samples $\pi$ per example.
Under shuffle, the predicted index $\hat{i}$ is mapped back to the original candidate index via $\pi$.

\section{KV Reuse for Multi-Candidate Judging}
Figure~\ref{fig:method} summarizes the judge-side cache construction strategies compared in this work. 
We consider judge-centric inference where a judge processes $N$ candidate responses within a single context.
We use $\mathcal{K}(\cdot)$ to denote the KV cache obtained by encoding a token sequence, and $\mathcal{K}(s)[u]$ to denote the KV entries in $\mathcal{K}(s)$ corresponding to a subsequence $u\subseteq s$.
For each candidate response $y_i$, the execution agent has already computed and cached its KV entries under its execution-time context $S_i^{\text{exec}}$:
\begin{equation}
\mathcal{C}_i \;=\; \mathcal{K}\!\left(S_i^{\text{exec}}\right)[y_i].
\label{eq:exec_chunk}
\end{equation}

We use $\oplus$ to denote concatenation of KV-cache chunks in the displayed candidate order (i.e., stitching cache segments into a single judge cache).
Let $o_i$ be the starting \emph{token position} (index) of candidate $y_i$ in the judge input, and let $\mathcal{C}_i^{\rightarrow o_i}$ denote the position-aligned version of $\mathcal{C}_i$ (e.g., via RoPE~\cite{su2024roformer}  re-indexing) so that it is valid starting at position $o_i$ in the judge sequence. For a KV cache reuse method $m$, we denote by $\mathcal{C}^{J}_{m}$ the assembled KV cache segment corresponding to the concatenation of the $N$ candidate responses in the judge prompt .

\paragraph{Dense Prefill.}
Let $S^{J}=(p_J,x,y_1,\dots,y_N)$ denote the full judge input sequence.
Dense prefill recomputes the judge KV cache from scratch as $\mathcal{K}(S^{J})$.
In particular, the KV chunk for candidate $y_i$ under dense judging is
\begin{equation}
\mathcal{C}_i^{\text{dense}} \;=\; \mathcal{K}\!\left(S^{J}\right)[y_i],
\label{eq:dense_chunk}
\end{equation}
which differs from the execution-time chunk $\mathcal{C}_i$ due to the different conditioning context.

Then the full judge cache can be written as
\begin{equation}
\mathcal{C}^{J}_{\text{dense}}
=
\mathcal{C}^{\text{dense}}_1
\oplus
\mathcal{C}^{\text{dense}}_2
\oplus
\cdots
\oplus
\mathcal{C}^{\text{dense}}_N.
\label{eq:dense_full_focus}
\end{equation}

\paragraph{Na\"ive Reuse (Position-Only Reuse).}
Na\"ive reuse stitches cached response chunks into the judge context by position alignment and concatenation:
\begin{equation}
\mathcal{C}^{J}_{\text{naive}}
\;=\;
\mathcal{C}_1^{\rightarrow o_1}
\oplus
\mathcal{C}_2^{\rightarrow o_2}
\oplus
\cdots
\oplus
\mathcal{C}_N^{\rightarrow o_N}.
\label{eq:naive_stitch_infix}
\end{equation}
This approach reuses execution-side KV as-is and does not account for prefix-dependent KV deviations induced by the judge-side context. This position-only stitching strategy has been used in prior KV reuse settings, e.g., in RAG-style cache reuse~\cite{turborag} and in multi-agent interaction setups that reuse cached internal states~\cite{zou2025latent}.

\paragraph{KVCOMM (Anchor-based Offset Correction).}
KVCOMM improves reuse by adding an anchor-retrieved correction to each reused chunk~\cite{KVCOMM}.
For candidate $i$, KVCOMM constructs a matching view $v_i$ under the current judge-side prefix and retrieves a correction
\begin{equation}
\widehat{\Delta}_i = \texttt{RetrieveOffset}(v_i;\mathcal{A}^{(i)}),
\end{equation}
where $\mathcal{A}^{(i)}$ is the agent-specific anchor pool used by KVCOMM.
It then corrects the aligned chunk as
\begin{equation}
\widetilde{\mathcal{C}}_i^{J} \;=\; \mathcal{C}_i^{\rightarrow o_i} + \widehat{\Delta}_i,
\label{eq:kvcomm_correct_infix}
\end{equation}
and assembles the judge cache by concatenation:
\begin{equation}
\mathcal{C}^{J}_{\text{kvcomm}}
\;=\;
\widetilde{\mathcal{C}}_1^{J}
\oplus
\widetilde{\mathcal{C}}_2^{J}
\oplus
\cdots
\oplus
\widetilde{\mathcal{C}}_N^{J}.
\end{equation}
We follow KVCOMM for anchor construction, matching, and correction estimation, and use its reliability criterion to fall back to dense computation when reuse is deemed unreliable.

\paragraph{PAL-KV (Pooled-Anchor Lookup).}
PAL-KV (Pooled-Anchor Lookup KV reuse) is a minimal modification of KVCOMM that changes only the retrieval scope of anchors.
While KVCOMM retrieves $\widehat{\Delta}_i$ from an agent-specific pool $\mathcal{A}^{(i)}$, PAL-KV retrieves from the union of all pools:
\begin{equation}
\widehat{\Delta}_i^{\text{pal}} = \texttt{RetrieveOffset}\!\left(v_i;\bigcup_{j=1}^{N}\mathcal{A}^{(j)}\right).
\label{eq:pal_retrieve_infix}
\end{equation}
and applies
\begin{equation}
\widetilde{\mathcal{C}}_{i,\text{pal}}^{J} = \mathcal{C}_i^{\rightarrow o_i} + \widehat{\Delta}_i^{\text{pal}}.
\end{equation}
The assembled judge cache is
\begin{equation}
\mathcal{C}^{J}_{\text{pal}}
=
\widetilde{\mathcal{C}}_{1,\text{pal}}^{J}
\oplus
\widetilde{\mathcal{C}}_{2,\text{pal}}^{J}
\oplus
\cdots
\oplus
\widetilde{\mathcal{C}}_{N,\text{pal}}^{J}.
\label{eq:pal_full}
\end{equation}

Thus, PAL-KV modifies \emph{which} correction is selected, while keeping the reuse and correction mechanism identical to KVCOMM.

\section{Experiments}
\subsection{Experimental Setup}

\paragraph{Multi-candidate generation.}
We use the two candidate-generation regimes in Sec.~\ref{sec:task}, and generate $N{=}4$ candidates per example.
To isolate judge-side effects, we disable execution-side reuse and \emph{fix the candidate set}: for each example we generate the candidates once with dense prefill and reuse the identical candidate texts for all judge-side methods.
We evaluate both \texttt{no-shuffle} and \texttt{shuffle} candidate orders to quantify ordering effects on judge-side reuse.

% \paragraph{Multi-candidate generation.}
% We instantiate judge-centric inference using the two candidate generation regimes introduced in Sec.~\ref{sec:task}:
% \texttt{Progressive refinement}, where candidates are generated sequentially and later candidates condition on earlier ones, and
% \texttt{Parallel exploration}, where candidates are generated independently from similar base contexts.
% To isolate judge-side effects, we disable execution-side reuse and fix the candidate set: for each example we generate the same $N=4$ candidates once (with dense prefill) and reuse these identical candidates for all judge-side methods;
% To encourage diverse candidate outputs and provide comparable rationales for the judge, execution agents use few-shot prompting~\cite{brown2020language} and chain-of-thought reasoning~\cite{wei2022chain}, producing both a final answer and an explicit rationale.
% We enforce a structured output format~\cite{shorten2024structuredrag} for both execution agents and the judge, enabling reliable extraction of the final answer and the selected candidate index.
% To control for potential ordering effects, we evaluate both a fixed candidate presentation order (\texttt{no-shuffle}) and a randomized order sampled per example (\texttt{shuffle}).
\paragraph{Models and implementation.}
% To precisely analyze KV-cache reuse behaviors, we evaluate open-source models under a unified inference stack based on HuggingFace Transformers.
% Unless otherwise specified, all agents (execution and judge) use \texttt{Llama-3.2-3B-Instruct}~\cite{meta2024llama}.
% \zhenglin{add anchor each agent 5}

Our primary experiments use \texttt{Llama-3.2-3B-Instruct}~\cite{meta2024llama} for both execution agents and the judge, while ablations consider models from the Llama-3.1/3.2~\cite{grattafiori2024llama} and Qwen-2.5~\cite{yang2024qwen2} families across a range of sizes (3B--14B). To induce candidate diversity while keeping the judge deterministic, execution agents use temperature $0.2$ and the judge uses temperature $0$. For KVCOMM and PAL-KV, we use an anchor pool of size $|\mathcal{A}^{(i)}|=5$ per agent.

% We evaluate open-source models using a unified HuggingFace Transformers inference stack.
% Unless otherwise specified, all agents use \texttt{Llama-3.2-3B-Instruct}~\cite{meta2024llama}.
\subsection{Benchmarks}
We evaluate judge-centric KV reuse on three representative benchmarks spanning reasoning, programming, and knowledge-intensive domains:
\textbf{GSM8K.} A grade-school math reasoning benchmark~\cite{cobbe2021training};
\textbf{HumanEval.} A code generation benchmark consisting of programming tasks with unit tests~\cite{chen2021evaluating}.
\textbf{MMLU.} A multi-domain knowledge benchmark covering diverse subjects~\cite{hendrycks2021measuring}.

% \paragraph{GSM8K.}
% A grade-school math reasoning benchmark~\cite{cobbe2021training}.
% % Execution agents generate step-by-step solutions, and the judge selects the best candidate based on correctness and reasoning quality.

% \paragraph{HumanEval.}
% A code generation benchmark consisting of programming tasks with unit tests~\cite{chen2021evaluating}.
% % Execution agents propose different implementations, and the judge selects the candidate that is most likely to satisfy the specification.

% \paragraph{MMLU.}
% A multi-domain knowledge benchmark covering diverse subjects~\cite{hendrycks2021measuring}.
% % Execution agents provide competing answers (often with brief justification), and the judge selects the most plausible one, requiring factual and conceptual comparison across candidates.

\subsection{Evaluation Metrics}

% \paragraph{Task Acc.}
% We report \textbf{Task Acc} as the primary end-task metric of the judge's final answer.
% For each benchmark, \textbf{Task Acc} corresponds to its standard evaluation metric: exact-match accuracy for GSM8K and MMLU, and Pass@1 for HumanEval.
% For simplicity, we denote all of them as \emph{Acc.} in tables.

% \paragraph{Judge Consistency Rate.}
% We report \textbf{Judge Consistency Rate} (JCR), defined as the percentage of examples where a KV reuse method leads the judge to select the \emph{same candidate index} as \emph{dense-prefill judging} on the \emph{same fixed candidate set}.
% JCR measures the \emph{invariance of the judge's selection/attribution decision} under judge-side KV reuse.
% For \texttt{shuffle}, we sample one candidate permutation per example and reuse the \emph{same} permutation for both dense and reuse judge runs; selected indices are mapped back to the original candidate ids before computing JCR.

% \paragraph{Reuse Rate.}
% We report \textbf{Reuse Rate} (Reuse), defined on the \emph{judge side} as the fraction of candidate blocks whose KV chunks are assembled via reuse (after any reliability checks and fallbacks) rather than recomputed densely under the judge context.
% Reuse is measured \emph{only} over the $N$ candidate responses $y_{1:N}$ in the judge input, and does not include the shared judge prefix (instruction/question) or the judge's own generation tokens.
% In all main experiments, we \emph{disable execution-side reuse} and keep the candidate set identical across methods; thus, Reuse reflects judge-side cache assembly only.

\paragraph{Task Acc.} 
The primary metric for the judge's final answer: exact-match accuracy for GSM8K and MMLU, and Pass@1 for HumanEval (also denoted as \emph{Acc.} in tables).

\paragraph{Judge Consistency Rate (JCR).} 
The percentage of examples where KV reuse and dense-prefill judging select the same candidate \emph{given the same candidate ordering}.  JCR measures the decision invariance under KV reuse. For \texttt{shuffle}, we use identical permutations for both runs and map indices back to original IDs.

\paragraph{Reuse Rate (Reuse).} 
The fraction of candidate blocks ($y_{1:N}$) assembled via reuse versus dense recomputation. It excludes the shared judge prefix and output tokens. We disable execution-side reuse to ensure an identical candidate set across methods.

% To induce candidate diversity while keeping the judge deterministic, execution agents use temperature $0.2$ and the judge uses temperature $0$. For KVCOMM and PAL-KV, we use an anchor pool of size $|\mathcal{A}^{(i)}|=5$ per agent.

\begin{table*}[htbp]
\centering
\caption{Judge-side KV cache reuse results on three benchmarks under different settings.
Bold denotes the best Acc., and the JCR column additionally reports the performance drop under \texttt{shuffle} relative to \texttt{no-shuffle}.}
\label{tab:kv_reuse_results}
\resizebox{\linewidth}{!}{
\begin{tabular}{ll l| ccc ccc ccc}
\toprule
\multirow{2}{*}{\textbf{Settings}} & \multirow{2}{*}{\textbf{Shuffle}} & \multirow{2}{*}{\textbf{Method}}
& \multicolumn{3}{c}{\textbf{MMLU}}
& \multicolumn{3}{c}{\textbf{GSM8K}}
& \multicolumn{3}{c}{\textbf{HumanEval}} \\
\cmidrule(lr){4-6} \cmidrule(lr){7-9} \cmidrule(lr){10-12}
& & 
& \textbf{Acc.} & \textbf{JCR\phantom{\scriptsize (-00.00)}} & \textbf{Reuse}
& \textbf{Acc.} & \textbf{JCR\phantom{\scriptsize (-00.00)}} & \textbf{Reuse}
& \textbf{Acc.} & \textbf{JCR\phantom{\scriptsize (-00.00)}} & \textbf{Reuse} \\
\midrule

\multirow{8}{*}{\textbf{\makecell[c]{Progressive\\Refinement}}}
& \multirow{4}{*}{\textbf{No}}
& \textbf{Dense Prefill} & 45.09 & 100.00\phantom{\scriptsize (-00.00)} & 00.00 & \textbf{71.34} & 100.00\phantom{\scriptsize (-00.00)} & 00.00 & 33.54 & 100.00\phantom{\scriptsize (-00.00)} & 00.00 \\
& & \textbf{Na\"ive Reuse} & \textbf{49.02} & 31.37\phantom{\scriptsize (-00.00)} & 100.00 & 11.83 & 50.87\phantom{\scriptsize (-00.00)} & 100.00 & 45.96 & 64.60\phantom{\scriptsize (-00.00)} & 100.00 \\
& & \textbf{KVCOMM} & 39.22 & 66.67\phantom{\scriptsize (-00.00)} & 32.55 & 64.14 & 87.91\phantom{\scriptsize (-00.00)} & 29.93 & \textbf{51.55} & 57.14\phantom{\scriptsize (-00.00)} & 48.32 \\
& & \textbf{PAL-KV} & 42.48 & 69.28\phantom{\scriptsize (-00.00)} & 32.55 & 64.59 & 89.23\phantom{\scriptsize (-00.00)} & 29.93 & 49.69 & 66.46\phantom{\scriptsize (-00.00)} & 48.32 \\

\cmidrule(lr){2-12}

& \multirow{4}{*}{\textbf{Yes}}
& \textbf{Dense Prefill} & 44.44 & 100.00\scriptsize\textcolor{PaperGreen}{(+00.00)} & 00.00 & \textbf{70.51} & 100.00\scriptsize\textcolor{PaperGreen}{(+00.00)} & 00.00 & 31.68 & 100.00\scriptsize\textcolor{PaperGreen}{(+00.00)} & 00.00 \\
& & \textbf{Na\"ive Reuse} & 39.86 & 27.45\scriptsize\textcolor{PaperRed}{(-03.92)} & 100.00 & 16.98 & 19.88\scriptsize\textcolor{PaperRed}{(-30.99)} & 100.00 & 49.06 & 23.60\scriptsize\textcolor{PaperRed}{(-41.00)} & 100.00 \\
& & \textbf{KVCOMM} & 39.87 & 39.87\scriptsize\textcolor{PaperRed}{(-26.80)} & 32.29 & 63.84 & 51.75\scriptsize\textcolor{PaperRed}{(-36.16)} & 31.86 & \textbf{49.69} & 21.74\scriptsize\textcolor{PaperRed}{(-35.40)} & 44.84 \\
& & \textbf{PAL-KV} & \textbf{47.06} & 31.37\scriptsize\textcolor{PaperRed}{(-37.91)} & 32.29 & 62.53 & 51.27\scriptsize\textcolor{PaperRed}{(-37.96)} & 31.86 & 49.07 & 24.22\scriptsize\textcolor{PaperRed}{(-42.24)} & 44.84 \\

\midrule

\multirow{8}{*}{\textbf{\makecell[c]{Parallel\\Exploration}}}
& \multirow{4}{*}{\textbf{No}}
& \textbf{Dense Prefill} & 49.02 & 100.00\phantom{\scriptsize (-00.00)} & 00.00 & \textbf{67.32} & 100.00\phantom{\scriptsize (-00.00)} & 00.00 & 28.57 & 100.00\phantom{\scriptsize (-00.00)} & 00.00 \\
& & \textbf{Na\"ive Reuse} & \textbf{61.44} & 26.14\phantom{\scriptsize (-00.00)} & 100.00 & 20.34 & 23.93\phantom{\scriptsize (-00.00)} & 100.00 & \textbf{40.99} & 39.75\phantom{\scriptsize (-00.00)} & 100.00 \\
& & \textbf{KVCOMM} & 47.06 & 58.16\phantom{\scriptsize (-00.00)} & 44.84 & 39.63 & 59.19\phantom{\scriptsize (-00.00)} & 63.12 & 31.68 & 61.49\phantom{\scriptsize (-00.00)} & 66.96 \\
& & \textbf{PAL-KV} & 45.10 & 62.75\phantom{\scriptsize (-00.00)} & 44.84 & 41.17 & 59.55\phantom{\scriptsize (-00.00)} & 63.12 & 32.30 & 64.60\phantom{\scriptsize (-00.00)} & 66.96 \\

\cmidrule(lr){2-12}

& \multirow{4}{*}{\textbf{Yes}}
& \textbf{Dense Prefill} & \textbf{51.63} & 100.00\scriptsize\textcolor{PaperGreen}{(+00.00)} & 00.00 & \textbf{70.28} & 100.00\scriptsize\textcolor{PaperGreen}{(+00.00)} & 00.00 & 25.63 & 100.00\scriptsize\textcolor{PaperGreen}{(+00.00)} & 00.00 \\
& & \textbf{Na\"ive Reuse} & 36.60 & 24.18\scriptsize\textcolor{PaperRed}{(-1.96)} & 100.00 & 19.86 & 24.92\scriptsize\textcolor{PaperGreen}{(+0.99)} & 100.00 & \textbf{50.31} & 27.95\scriptsize\textcolor{PaperRed}{(-11.80)} & 100.00 \\
& & \textbf{KVCOMM} & 43.14 & 46.41\scriptsize\textcolor{PaperRed}{(-11.75)} & 44.84 & 38.89 & 34.43\scriptsize\textcolor{PaperRed}{(-24.76)} & 63.21 & 37.27 & 32.92\scriptsize\textcolor{PaperRed}{(-28.57)} & 66.96 \\
& & \textbf{PAL-KV} & 43.14 & 45.75\scriptsize\textcolor{PaperRed}{(-17.00)} & 44.84 & 39.95 & 35.32\scriptsize\textcolor{PaperRed}{(-24.23)} & 63.21 & 36.02 & 36.65\scriptsize\textcolor{PaperRed}{(-27.95)} & 66.96 \\

\bottomrule
\end{tabular}
}
\end{table*}

\section{Results and Analysis}

\subsection{Main Results}
\label{sec:main_results}
\paragraph{Overview: judge decisions are not invariant under KV reuse.}
Table~\ref{tab:kv_reuse_results} reports task accuracy (Acc.), Judge Consistency Rate (JCR), and the judge-side candidate Reuse Rate (Reuse) across all benchmarks.
A central observation is that \textbf{KV reuse can substantially change which candidate the judge selects, even when end-task accuracy remains comparable}.
This decision non-invariance appears consistently under both multi-agent candidate-generation regimes (\texttt{Progressive Refinement} and \texttt{Parallel Exploration}), suggesting that the phenomenon is not tied to a particular collaboration pattern.
Overall, these results expose a failure mode that would be missed if one monitors task metrics alone: KV reuse may preserve outcomes while silently perturbing the judge's underlying selection behavior.

\paragraph{Accuracy can mask instability: reuse perturbs selection beyond what Acc reveals.}
Under dense prefill, changing the candidate order (\texttt{shuffle} vs.\ \texttt{no-shuffle}) often affects Acc. only moderately.
In contrast, reuse-based methods frequently exhibit \textbf{low JCR} even when Acc. remains close to the dense-prefill baseline, indicating that the judge's cross-candidate comparison is highly sensitive to judge-side reuse.
This decoupling is critical for judge-centric pipelines where downstream properties---such as attribution, explanation, and auditability---depend on \emph{which} candidate is chosen, not only whether the final answer is correct.
Appendix~\ref{difficulty} further analyzes this phenomenon and shows that Acc. and JCR are weakly correlated.

\paragraph{Candidate-order perturbation (shuffle) amplifies inconsistency.}
A striking pattern is that \texttt{shuffle} sharply reduces JCR for \emph{all} reuse-based methods.
This indicates that judge-side reuse is strongly \emph{layout-dependent}: the effective prefix for each candidate block is determined by the entire preceding candidate configuration in the judge prompt.
When the order changes, reuse approximations become much less reliable, leading to large drops in JCR even when Acc. changes little.

\paragraph{PAL-KV as a probe: agent identity is not the dominant bottleneck.}
We use PAL-KV as a controlled variant to test whether judge inconsistency is mainly caused by \emph{agent-specific} offset distributions assumed by KVCOMM.
By pooling anchors across agents, PAL-KV relaxes the agent-identity constraint in offset retrieval.
Empirically, pooled retrieval yields \emph{small but consistent} gains in the fixed-layout setting (\texttt{no-shuffle}), suggesting that some offset patterns transfer across agents when the presentation structure is stable.
However, under \texttt{shuffle}, PAL-KV does not mitigate the large JCR drop and behaves similarly to KVCOMM; other probes show the same limitation (Appendix~\ref{Slot}).
Together, these results suggest that \textbf{the dominant driver of judge-centric failures is the changing cross-candidate context/layout, rather than agent identity per se}. In the next subsection, we provide direct evidence that preserving \emph{cross-candidate interaction} is crucial for maintaining dense-prefill judge decisions.

\begin{table}[t]
  \centering
  \caption{Effect of masking cross-candidate attention under dense prefill.}
  \label{tab:masking_setting}
  \resizebox{\linewidth}{!}{
  \begin{tabular}{llcccc}
    \toprule
    \multirow{2}{*}{\textbf{Settings}} & \multirow{2}{*}{\textbf{Shuffle}} & \multicolumn{2}{c}{\textbf{Acc. (\%)}} & \multicolumn{2}{c}{\textbf{JCR (\%)}} \\
 \cmidrule(lr){3-4} \cmidrule(lr){5-6}
    &  & \textbf{Original} & \textbf{Masked} & \textbf{Original} & \textbf{Masked} \\
    \midrule
    \multirow{2}{*}{\textbf{\makecell[c]{Progressive\\Refinement}}} 
        & \textbf{No} & 45.09 & 43.79 & 100.00 & 28.76 \\
        & \textbf{Yes}   & 44.44 & 45.10 & 100.00 & 32.03 \\
    \addlinespace
    \multirow{2}{*}{\textbf{\makecell[c]{Parallel\\Exploration}}}
        & \textbf{No} & 49.02 & 48.37 & 100.00 & 22.22 \\
        & \textbf{Yes}   & 54.25 & 47.71 & 100.00 & 31.37 \\
    \bottomrule
  \end{tabular}
  }
\end{table}
\begin{figure}[t]
    \centering
    \includegraphics[width=1\linewidth]{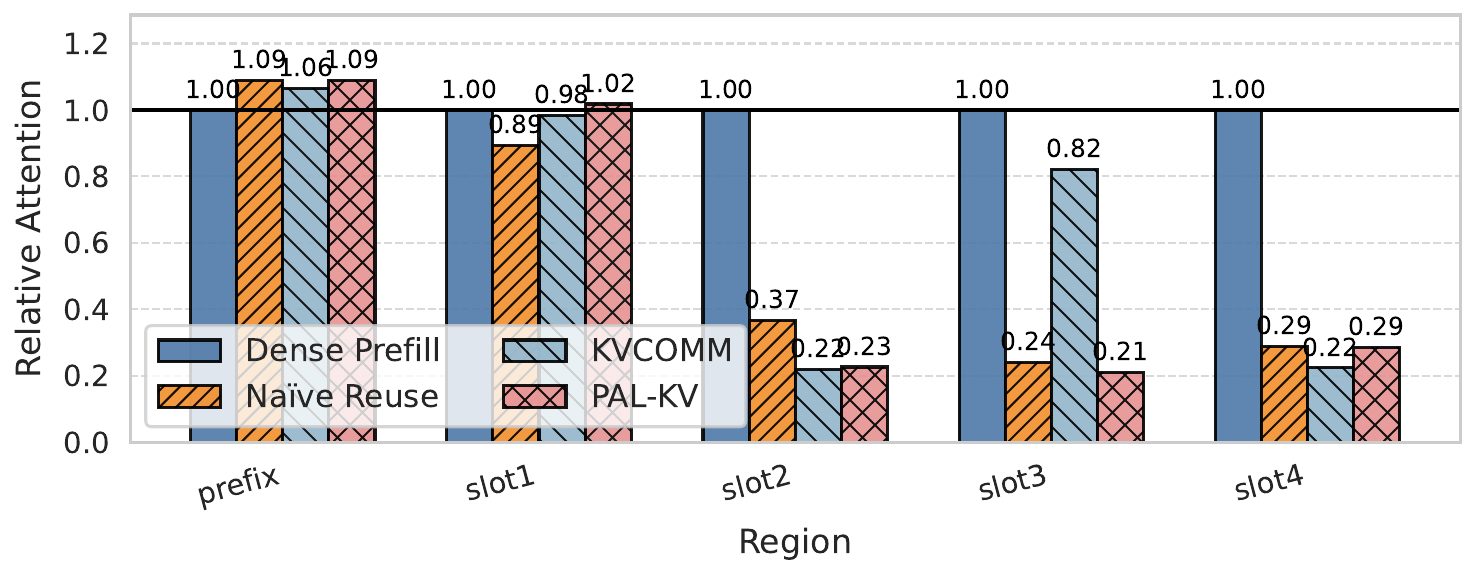}
    \caption{Relative attention mass over regions (prefix and candidate slots) under different KV reuse methods.}
    \label{fig:attn}
\end{figure}
\subsection{Why Does Judge-Centric Reuse Fail? Interaction Diagnostics}
\label{sec:diagnostics}

\paragraph{Attention diagnostics: reuse weakens cross-candidate attention, especially for later slots.}
To understand why judge selections change under reuse, we analyze attention patterns during the judge's first-token generation.
Figure~\ref{fig:attn} shows a consistent trend: attention over the shared prefix and the first candidate block is relatively similar across methods, but \textbf{attention to later candidate blocks is substantially weaker and more erratic under reuse-based methods} than under dense prefill. Representative attention maps for individual methods are provided in Appendix~\ref{attention}.
This is consistent with the interpretation that reuse perturbs the fine-grained cross-candidate interactions needed for joint comparison, causing the judge to under-attend to late-arriving evidence.

\paragraph{Masking ablation: cross-candidate interaction is necessary beyond judge-style formatting.}
A natural alternative explanation is that reuse fails simply because execution-side KV caches are ``incompatible'' with the judge prompt.
To isolate the role of \emph{interaction} from \emph{format}, we conduct a masking ablation where candidates are placed in the judge prompt but cross-candidate attention is explicitly blocked (candidate $i$ cannot attend to candidates $1{:}i{-}1$).
As shown in Table~\ref{tab:masking_setting}, masking has only a limited effect on \emph{Acc} but causes a dramatic collapse in \emph{JCR}, approaching random-choice behavior for $N{=}4$.
This provides direct evidence that \textbf{judge-centric inference critically relies on cross-candidate interaction}, and that removing such interaction---even under the same judge-style prompt---destroys decision invariance relative to dense prefill.

\begin{figure}[t]
    \centering

    \begin{subfigure}[t]{\linewidth}
        \centering
        \includegraphics[width=\linewidth]{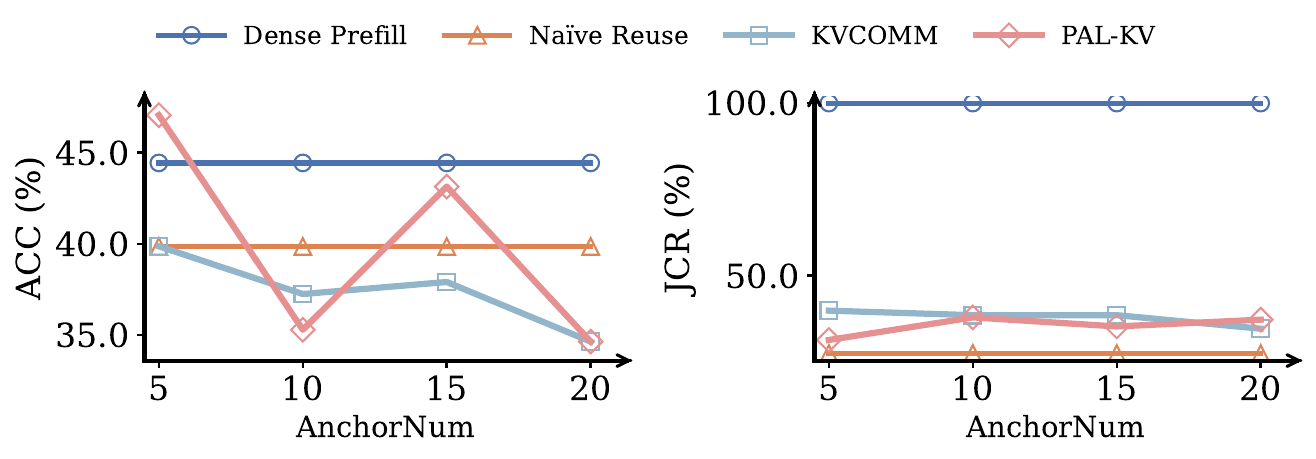}
        \caption{Effect of anchor pool size.}
        \label{fig:ab_anchor}
    \end{subfigure}
    
    \begin{subfigure}[t]{\linewidth}
        \centering
        \includegraphics[width=\linewidth]{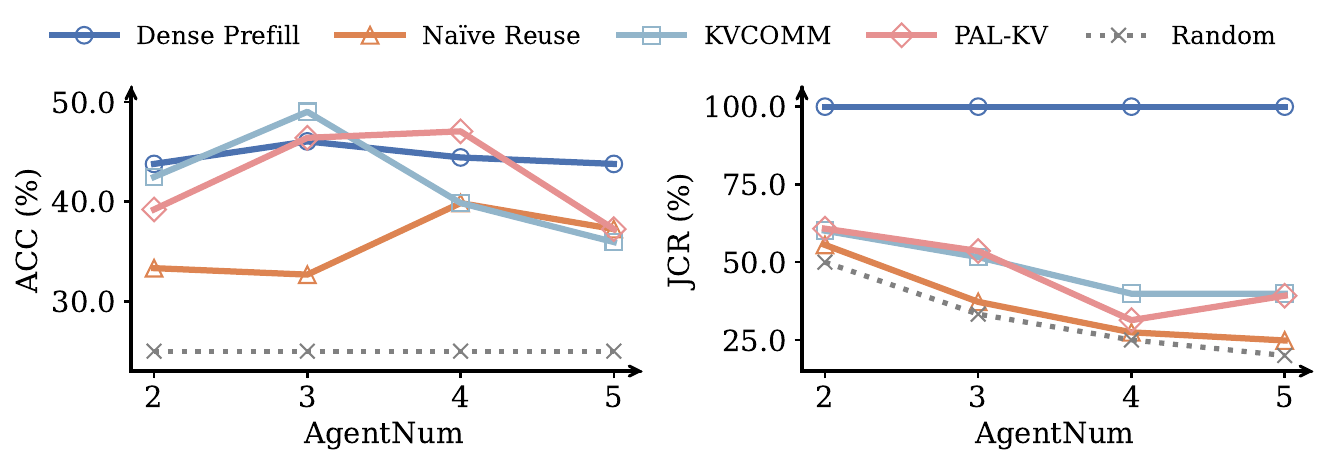}
        \caption{Effect of the number of candidates (agents).}
        \label{fig:ab_agent}
    \end{subfigure}

    \begin{subfigure}[t]{\linewidth}
        \centering
        \includegraphics[width=\linewidth]{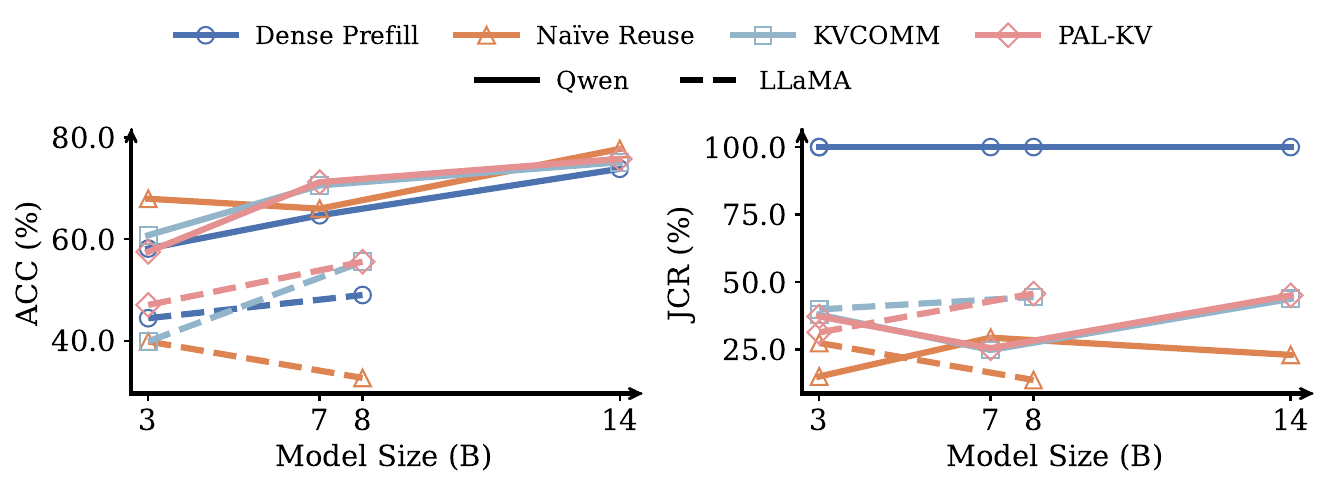}
        \caption{Effect of model family and size.}
        \label{fig:ab_size}
    \end{subfigure}

    \caption{Ablations under \texttt{shuffle}: judge-side decision non-invariance persists. Varying anchor pool size, candidate count, or model size does not reliably restore JCR for KV reuse methods.}
    \label{fig:ab_all}
\end{figure}

\subsection{Ablation Studies}
\label{sec:ablation}

% Unless otherwise specified, all ablations are conducted on \textbf{MMLU under Progressive Refinement with shuffled candidate order}, which is the most challenging judge-centric setting.

Unless otherwise specified, all ablations focus on the \texttt{shuffle} regime, which is particularly challenging for judge-side reuse due to order-induced layout changes; we use \textbf{MMLU under Progressive Refinement with shuffled candidate order} as a representative testbed.

\paragraph{Anchor pool size is not the bottleneck under shuffle.}
A plausible hypothesis is that poor anchor matches cause offset correction to fail.
We therefore increase the anchor pool size (Figure~\ref{fig:ab_anchor}); as expected, a larger pool increases the Reuse Rate (Appendix Table~\ref{tab:anchor_num}). 
Nevertheless, JCR under \texttt{shuffle} stays low and changes little as the pool grows, suggesting that the failure is \textbf{not} driven by insufficient anchor coverage.
Instead, when the preceding candidate configuration changes, anchor similarity becomes an unreliable proxy for preserving the interaction effects needed for judging.

\paragraph{More candidates can improve Acc but make decision invariance harder.}
We vary the number of candidates $N$ (Figure~\ref{fig:ab_agent}).
Increasing $N$ often improves \emph{Acc} by providing stronger candidates, but \textbf{JCR decreases monotonically} across reuse-based methods.
This highlights a key tension in judge-centric systems: adding candidates improves solution quality but expands the space of cross-candidate interactions that reuse must preserve, making invariant selection increasingly fragile.

\paragraph{Scaling model size improves Acc but does not reliably restore JCR.}
We evaluate different model families and sizes (Figure~\ref{fig:ab_size}).
While larger models tend to improve \emph{Acc}, \textbf{JCR does not show a corresponding improvement} and remains unstable under reuse.
This indicates that judge inconsistency is not merely a capacity issue; rather, it reflects sensitivity to interaction patterns that current reuse schemes fail to preserve.

% \begin{table}[t]
%   \centering
%   \caption{Effect of masking under different settings.}
%   \label{tab:masking_setting}
%   \resizebox{\linewidth}{!}{
%   \begin{tabular}{llcccc}
%     \toprule
%     \multirow{2}{*}{Settings} & \multirow{2}{*}{Shuffle} & \multicolumn{2}{c}{Acc. (\%)} & \multicolumn{2}{c}{JCR (\%)} \\
%  \cmidrule(lr){3-4} \cmidrule(lr){5-6}
%     &  & Original & Masked & Original & Masked \\
%     \midrule
%     \multirow{2}{*}{\makecell[c]{Progressive\\Refinement}} 
%         & No & 45.09 & 43.79 & 100.00 & 28.76 \\
%         & Yes   & 44.44 & 45.10 & 100.00 & 32.03 \\
%     \addlinespace
%     \multirow{2}{*}{\makecell[c]{Parallel\\Exploration}}
%         & No & 49.02 & 48.37 & 100.00 & 22.22 \\
%         & Yes   & 54.25 & 47.71 & 100.00 & 31.37 \\
%     \bottomrule
%   \end{tabular}
%   }
% \end{table}

% \section{Discussion and Future Directions}
\begin{figure}[t]
    \centering
    \includegraphics[width=1\linewidth]{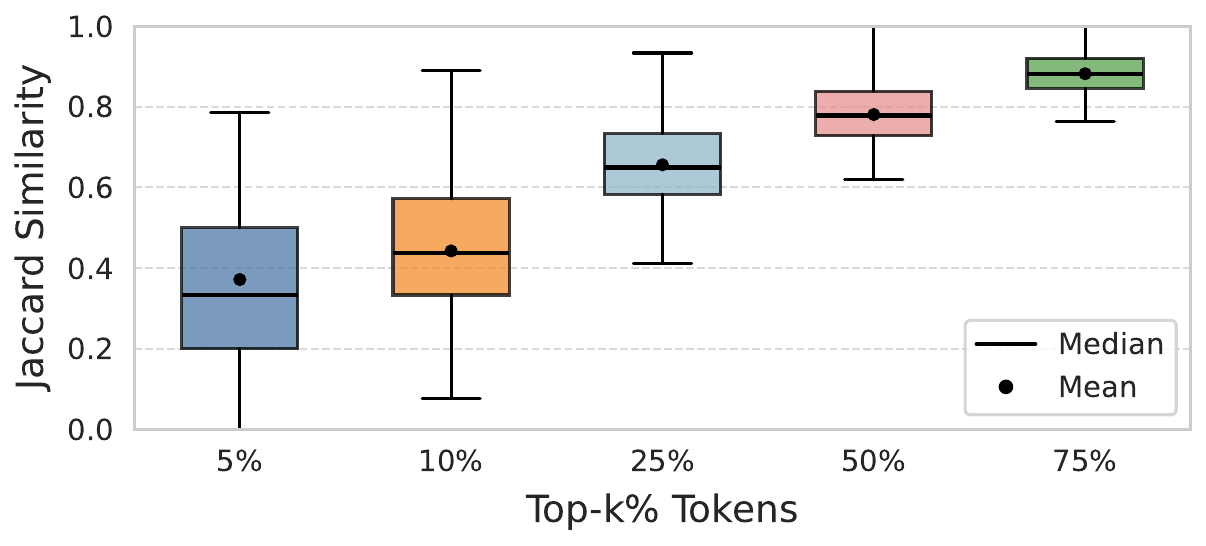}
    \caption{Jaccard similarity of selected Top-$k$\% tokens between the small and large models.}
    \label{fig:jaccard}
\end{figure}

% \begin{figure}[t]
%     \centering
%     \includegraphics[width=1\linewidth]{tsne.pdf}
%     \caption{}
%     \label{fig:tsne}
% \end{figure}

\begin{figure}[t]
    \centering
    \includegraphics[width=1\linewidth]{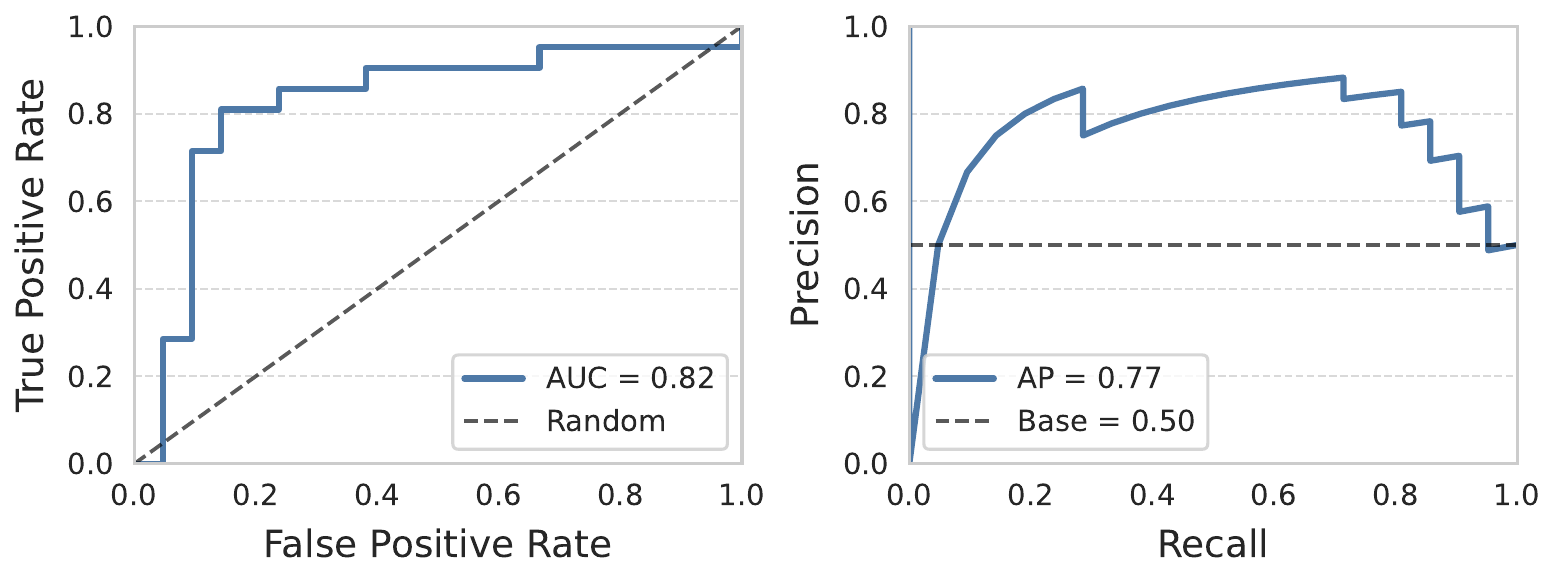}
    \caption{Detecting universally safe vs.\ unsafe instances for judge-side reuse (AUC $\approx 0.82$, AP $\approx 0.77$).}
    \label{fig:auc}
\end{figure}
\section{Discussion and Future Directions}
\label{sec:discussion}

Our study highlights a judge-centric failure mode of KV cache reuse: when a judge must \emph{jointly} compare multiple candidate blocks, reuse can substantially alter the selected candidate (low JCR), especially under unstable layouts such as \texttt{shuffle}.
Together with attention diagnostics and the masking ablation, the evidence points to a central mechanism: \textbf{judge-centric inference relies on fine-grained cross-candidate interactions that current reuse schemes do not reliably preserve}.
We discuss two preliminary directions suggested by this mechanism.

% \paragraph{Direction 1: Interaction-preserving reuse via selective context retention.}
\paragraph{From the perspective of interaction-preserving}
% A promising direction is to make judge-side reuse more faithful by preserving the interaction-relevant parts of the context.
% One hypothesis is that removing \emph{irrelevant} tokens from the KV cache (e.g., redundant rationale segments or low-salience prefix tokens) could reduce attention dilution and help the judge allocate attention more evenly across informative evidence from multiple candidates.\zhenglin{\cite{}add some ref...}
% Importantly, this direction does not reduce cross-candidate visibility; it aims to make cross-candidate comparison more focused under a limited KV budget.
% In a pilot analysis, we compare token selections induced by a small model (Llama-3.2-1B-Instruct) and a larger model (Llama-3.2-3B-Instruct) and compute the Jaccard similarity of selected token sets (Figure~\ref{fig:jaccard}).
% We observe substantial overlap that increases with the retention budget, suggesting that small--large cooperation may help identify judge-relevant tokens.
% However, whether such retention can reliably restore judge invariance under reuse remains open and likely requires interaction-aware objectives beyond token overlap.
A promising direction is to enhance judge-side reuse by selectively retaining interaction-relevant context. 
Removing low-salience tokens (e.g., redundant rationales) may mitigate attention dilution~\cite{shi2023large}, helping the judge focus on informative evidence across candidates.
Crucially, this aims to concentrate rather than restrict cross-candidate visibility.
In a pilot analysis, we find substantial overlap in token selections between Llama-3.2-1B and 3B models, with Jaccard similarity increasing alongside the retention budget (Figure~\ref{fig:jaccard}).
This suggests that small-to-large model cooperation~\cite{zhao2025smallkv,liu2025speculative} could help identify judge-relevant tokens.
However, fully restoring judge invariance likely requires more sophisticated, interaction-aware objectives beyond simple token-level overlap.
\begin{findings}
\textbf{\underline{Future Direction (i)}}: \textit{Interaction-preserving reuse via selective context retention.}
\end{findings}

% \paragraph{Direction 2: Risk-aware gating for judge-side reuse.}
% \paragraph{From the perspective of meta-reasoning}
% Given the sensitivity of judge decisions under reuse, another direction is to gate reuse with conservative fallbacks.
% \zhenglin{add some ref...}
% Rather than predicting which reuse strategy will succeed, a more robust goal is to identify instances that are \emph{universally safe} for reuse (i.e., unlikely to change the judge's selection) and use dense prefill otherwise.
% Using features derived from the candidate set (e.g., correctness signals and candidate similarity statistics), a simple classifier can distinguish ``universally safe'' vs.\ ``universally unsafe'' instances with AUC around 0.82 (Figure~\ref{fig:auc}), indicating that extreme cases are separable.
% However, predicting the success of a specific reuse strategy remains much harder (AUC $\approx 0.6$), suggesting that practical gating should prioritize conservative safety detection over fine-grained strategy selection.
% \begin{findings}
% \textbf{\underline{Future Direction (ii)}}: \textit{Risk-aware gating for judge-side reuse.}
% \end{findings}

\paragraph{From the perspective of meta-reasoning for risk-aware gating.}
Given the judge's sensitivity to KV reuse, a promising direction is to apply \emph{meta-reasoning}—deciding which inference strategy to employ based on the specific situation~\cite{yan2025position}.
While proactive segment prediction is effective in standard KV systems~\cite{pan2025kvflow}, judge-centric reuse requires conservative gating with reliable fallbacks. 
Rather than predicting the success of a specific strategy (AUC $\approx 0.6$), a more robust meta-goal is to identify instances that are \emph{universally safe} for reuse (i.e., unlikely to alter the judge's selection in all three KV Reuse Methods). 
Using features derived from the candidate set—such as correctness signals and pairwise similarities—a simple classifier can distinguish ``universally safe'' vs.\ ``unsafe'' instances with an AUC of 0.82 (Figure~\ref{fig:auc}). 
This suggests that future gating mechanisms should prioritize conservative safety detection over fine-grained strategy selection to ensure decision invariance.
\begin{findings}
\textbf{\underline{Future Direction (ii)}}: \textit{Meta-reasoning and risk-aware fallback gating for judge-side reuse.}
\end{findings}

\paragraph{Outlook.}
Overall, accelerating judge-centric inference likely requires methods that explicitly preserve cross-candidate interactions under variable layouts, potentially combining interaction-aware reuse, selective retention/compression, and conservative gating with principled fallbacks.
% \begin{findings}
% \textbf{\underline{Promising Considerations}}: \textit{Layout-robust context alignment, selective KV retention, and conservative reuse-side gating.}
% \end{findings}

\section{Conclusion}
Our results reveal that KV reuse in judge-centric tasks is not inherently behavior-preserving.
While task accuracy remains stable, JCR exposes a critical decoupling: judges frequently switch selections due to disrupted cross-candidate interactions, especially under variable layouts.
These findings establish JCR as a vital diagnostic and motivate future research into interaction-preserving and risk-aware KV acceleration.

% We show that KV cache reuse is not always behavior-preserving for \emph{judge-centric} multi-candidate inference: across GSM8K, MMLU, and HumanEval, accuracy can remain stable while the judge’s selected candidate becomes inconsistent with dense prefill, particularly under shuffling.
% JCR quantifies this decision non-invariance, and analyses point to disrupted cross-candidate interaction as the underlying mechanism, motivating interaction- and risk-aware judge-side acceleration.

% We studied KV cache reuse in \emph{judge-centric} multi-candidate inference, where an LLM judge must jointly compare several candidates within a single context.
% Across GSM8K, MMLU, and HumanEval, we find that reuse strategies effective for execution agents can substantially alter judge behavior: task accuracy may appear stable, yet the selected candidate becomes highly inconsistent with dense prefill.
% We capture this risk with Judge Consistency Rate (JCR) and show that perturbing candidate order further exacerbates inconsistency.
% Analyses of attention and a masking ablation indicate that preserving fine-grained cross-candidate interactions is crucial for maintaining dense-prefill decisions, and current reuse schemes often disrupt these interactions.
% These results reveal an overlooked failure mode of KV reuse and motivate risk-aware acceleration methods tailored to judge-centric inference.
\newpage
\section*{Limitations}

\paragraph{Judge biases and reliability are not the focus.}
LLM-based judges are known to exhibit intrinsic biases and instability (e.g., position preference and sensitivity to formatting).
This work does not aim to assess the judge's \emph{absolute} fairness, calibration, or stability, nor do we analyze whether KV reuse amplifies specific biases of LLM judges.
Instead, we focus on a narrower question: whether judge-side KV reuse is \emph{behavior-preserving} relative to dense prefill in multi-candidate \emph{joint} judging, i.e., whether reuse introduces decision non-invariance even when task accuracy appears stable.

\paragraph{Scope of reuse and system coverage.}
Our study focuses on \emph{judge-centric} KV reuse---reusing KV chunks of candidate blocks when constructing the judge's prefill cache---and uses dense prefill as the reference behavior.
We do not claim comprehensive end-to-end optimization for all components of a full multi-agent stack, nor do we report system-wide latency/throughput under different serving configurations.
Consequently, our reported Reuse Rate and analyses are intentionally centered on the judge stage rather than end-to-end efficiency.

\paragraph{Protocol and prompt dependence.}
Our findings are obtained under a specific judging protocol (joint multi-candidate selection with structured outputs) and candidate-generation settings (few-shot prompting and chain-of-thought style rationales).
Different judge prompts (e.g., pairwise ranking vs.\ joint selection), output constraints (selection-only vs.\ judge rewriting), or candidate formatting may change the sensitivity profile.
While we expect the core phenomenon---that joint judging relies on delicate cross-candidate interactions---to persist, quantitative results may vary with protocol choices.

\paragraph{Limited cross-model and cross-architecture reuse.}
We do not systematically evaluate KV reuse across \emph{different model families}, sizes, or architectures (e.g., heterogeneous agents and judges).
Such heterogeneous deployments are common in practice (e.g., a stronger judge supervising weaker generators, or mixing different model families), and the transferability of KV corrections in these settings remains unclear.
This limitation partly reflects the current research landscape: formally published cross-model KV reuse/sharing approaches appear relatively scarce, with representative examples targeting specific adaptation or sharing settings (e.g., MobiLoRA~\cite{mobilora} and DroidSpeak~\cite{droidspeak}), rather than broad cross-family, cross-size reuse.
Extending judge-centric reuse to heterogeneous model mixtures is an important direction for future work.

\paragraph{Model scale coverage.}
Due to computational constraints, our experiments focus on small-to-mid scale open-source LLMs (up to 14B parameters).
While we include a model-size ablation within this range, we do not claim that the magnitude of decision non-invariance (e.g., JCR drops under reuse) transfers unchanged to substantially larger models.
Evaluating judge-centric KV reuse on larger-scale LLMs (and potentially stronger judges supervising weaker generators) remains an important direction.
To this end, we are actively expanding our evaluation to include models with significantly larger parameter counts to further generalize the findings of this study.

\paragraph{Implementation stack and deployment variability.}
All experiments are conducted under an open-source inference stack (HuggingFace Transformers\footnote{\url{https://huggingface.co/docs/transformers}}) on a fixed GPU setup.
Different kernels, serving systems, attention implementations, or cache-management policies could affect absolute accuracy and the magnitude of reuse-induced perturbations.
We encourage validation under diverse deployment stacks.

\paragraph{Downstream safety and compute considerations.}
Our study highlights a failure mode where accuracy may mask decision instability; in high-stakes downstream workflows, such non-invariance can complicate accountability and auditing.
We do not provide application-specific deployment guidelines, and practitioners should perform domain-specific evaluation and safeguards.
Finally, while KV reuse is motivated by reducing redundant computation, we do not quantify end-to-end energy impacts; efficiency gains may be offset by increased scale of deployment.

\bibliography{custom}

\appendix
\newpage

\section{Additional Analyses and Extended Results}
\label{app:additional}

% \subsection{Additional Results for Na\"ive Reuse}
% \label{app:naive}
% \noindent
% We report additional observations for Na\"ive Reuse (position-only stitching) to complement the main results and clarify its failure mode in judge-centric joint comparison.
% % NOTE: You can keep the existing paragraph below (moved here) or shorten it.

\subsection{Na\"ive position-only reuse is brittle for judge-centric comparison}
Na\"ive Reuse reuses \emph{all} candidate blocks (Reuse\%=100\% by construction) via position alignment and stitching, without accounting for prefix-dependent deviations.
While this can be effective for some standard prefix-sharing workloads, it is highly unstable for judge-centric inference: it often produces very low JCR and can even catastrophically harm accuracy on reasoning benchmarks (e.g., severe drops on GSM8K), indicating that position-only stitching fails to preserve the cross-candidate interaction patterns required for reliable joint comparison.

\subsection{Judge Instability vs.\ Task Difficulty}
\label{difficulty}

\noindent
This section provides additional evidence that low judge consistency (low JCR) is not merely a byproduct of task difficulty.
Here we operationalize \emph{task difficulty} by \textbf{ACC Counts}: for each example, we count how many reuse-based methods (Na\"ive Reuse, KVCOMM, and PAL-KV) produce a correct final answer.
We measure decision non-invariance by \textbf{JCR Counts}: the number of reuse-based methods whose \emph{selected candidate} matches dense prefill (i.e., the per-method JCR indicator aggregated over the three reuse methods).

\begin{figure}[t]
    \centering
    \includegraphics[width=1\linewidth]{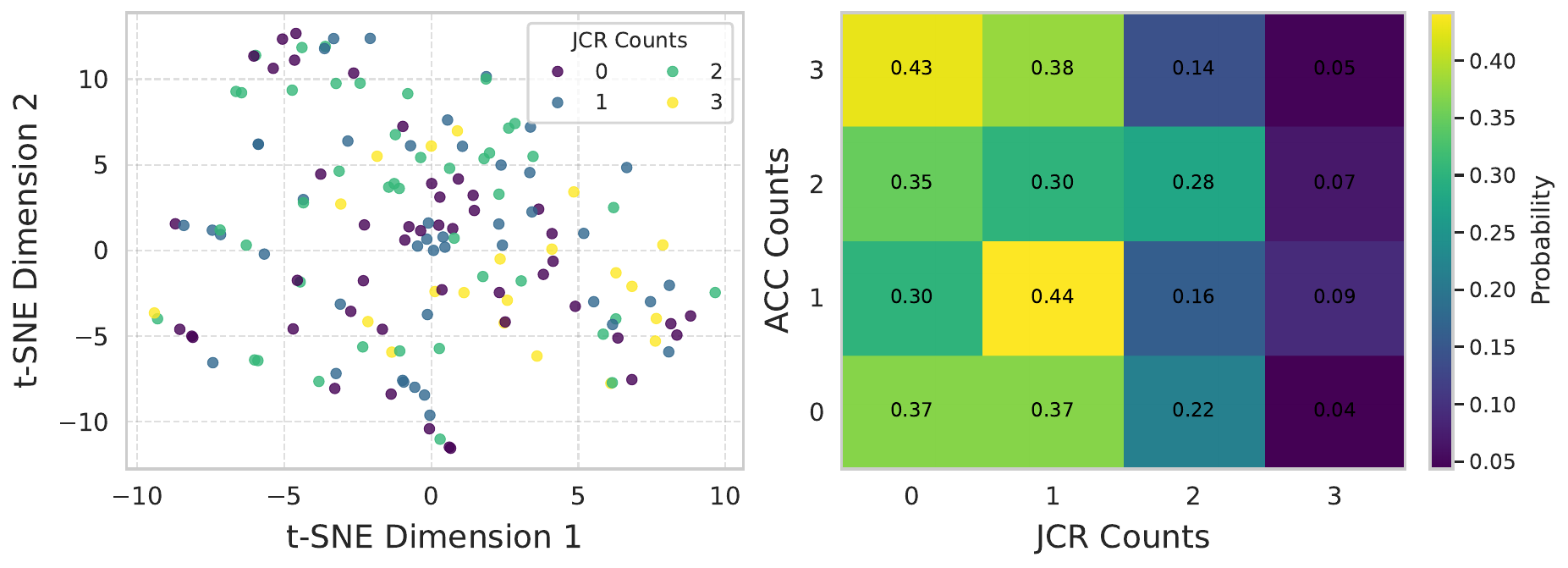}
    \caption{Task difficulty vs.\ judge decision non-invariance. \textbf{Left:} t-SNE of question embeddings, colored by \emph{JCR Counts} (number of reuse methods whose selection matches dense prefill). \textbf{Right:} joint distribution between \emph{ACC Counts} (number of reuse methods that answer correctly) and \emph{JCR Counts}.}
    \label{fig:tsne}
\end{figure}

\textbf{Embedding view.}
Figure~\ref{fig:tsne} (left) visualizes question embeddings projected to 2D with a supervised t-SNE, where points are colored by JCR Counts.
If judge non-invariance were primarily driven by question semantics (or by a small subset of intrinsically ``hard'' questions), we would expect instances with low JCR Counts to form separable regions.
Instead, instances with different JCR Counts are heavily intermixed, suggesting that \textbf{decision non-invariance is not well explained by the question embedding geometry} and is difficult to predict from semantics alone.

\textbf{Difficulty--consistency relationship.}
Figure~\ref{fig:tsne} (right) shows the empirical distribution of ACC Counts vs.\ JCR Counts.
While one might expect ``easy'' instances (high ACC Counts) to also yield stable judge selections (high JCR Counts), the heatmap shows only a weak association: even when multiple reuse methods answer correctly, the judge's selected candidate can still differ from dense prefill (non-trivial mass at high ACC Counts but low JCR Counts).
Conversely, low ACC Counts do not uniquely correspond to low JCR Counts either.
Overall, these results indicate that \textbf{task correctness and decision invariance are partially decoupled} in judge-centric inference, reinforcing the need for JCR-style diagnostics beyond end-task accuracy.

% \subsection{Judge Instability vs.\ Task Difficulty}
% \label{difficulty}

% \begin{figure}[h]
%     \centering
%     \includegraphics[width=1\linewidth]{tsne.pdf}
%     \caption{Additional analysis of task difficulty vs.\ decision non-invariance (JCR).}
%     \label{fig:tsne}
% \end{figure}

% \noindent
% This section provides additional evidence that decision non-invariance (low JCR) is not simply a byproduct of task difficulty.
% % TODO: Optionally add quantitative correlation or binning analysis.

% Low JCR under reuse does not simply reflect ``hard'' instances.
% As visualized in Figure~\ref{fig:tsne} (left), ``safe'' and ``unsafe'' instances are heavily intermixed in the embedding space, indicating that question semantics alone is not predictive of judge invariance.
% Furthermore, Figure~\ref{fig:tsne} (right) shows that task accuracy and judge consistency are only weakly correlated: easy instances (high \emph{Acc}) can still exhibit low JCR.
% This reinforces that \textbf{judge-centric inference is a distinct regime} where end-task metrics mask decision instability, necessitating JCR-style diagnostics.

\subsection{Anchor Pool Size and Effective Reuse}
\label{app:anchor_pool}
\noindent
As showed in Table~\ref{tab:anchor_num}, we vary the anchor pool size to test whether insufficient anchor coverage explains low JCR under \texttt{shuffle}.
As expected, a larger pool increases the effective Reuse Rate, but it does not reliably restore JCR.

\begin{table}[t]
  \centering
  \caption{Effect of anchor pool size on performance (additional results).}
  \label{tab:anchor_num}
  \resizebox{\linewidth}{!}{
  \begin{tabular}{lcccc}
    \toprule
    Method & AnchorNum & Acc. (\%) & JCR (\%) & Reuse (\%) \\
    \midrule
    \multirow{4}{*}{KVCOMM}
      & 5  & 39.87 & 39.87 & 32.29 \\
      & 10 & 37.25 & 38.56 & 47.58 \\
      & 15 & 37.91 & 38.56 & 52.55 \\
      & 20 & 34.64 & 34.64 & 57.25 \\
    \addlinespace
    \multirow{4}{*}{PAL-KV}
      & 5  & 47.06 & 31.37 & 32.29 \\
      & 10 & 35.29 & 37.91 & 47.58 \\
      & 15 & 43.14 & 35.29 & 52.55 \\
      & 20 & 34.64 & 37.25 & 57.25 \\
    \bottomrule
  \end{tabular}
  }
\end{table}

\subsection{Slot-Aligned Stabilization Baseline}
\label{Slot}
\noindent
We evaluate a slot-aligned variant that indexes reuse decisions by fixed candidate slots to test whether slot identity alone can stabilize reuse under \texttt{shuffle}.

One might hope that indexing offsets by fixed \emph{slots} in the judge prompt would stabilize reuse under shuffling.
However, Table~\ref{tab:slot_align} shows no consistent gain from a slot-aligned variant.
This suggests that true stability requires more than slot identity: for a candidate block, the \emph{entire preceding candidate configuration} (and thus the induced interaction structure) matters.
Under \texttt{shuffle}, this configuration space scales combinatorially, making naive stabilization impractical.

\begin{table}[t]
  \centering
  \caption{Slot-aligned stabilization does not reliably improve decision invariance (additional results).}
  \label{tab:slot_align}
  \resizebox{\linewidth}{!}{
  \begin{tabular}{llccc}
    \toprule
    Setting & Method & Acc. (\%) & JCR (\%) & Reuse (\%) \\
    \midrule
    \multirow{2}{*}{\makecell[c]{Progressive\\Refinement}}
      & KVCOMM     & 39.87 & 39.87 & 32.29 \\
      & Slot-Align & 43.79 & 36.84 & 26.14 \\
    \addlinespace
    \multirow{2}{*}{\makecell[c]{Parallel\\Exploration}}
      & KVCOMM     & 43.14 & 46.41 & 44.84 \\
      & Slot-Align & 43.14 & 45.10 & 44.44 \\
    \bottomrule
  \end{tabular}
  }
\end{table}

\begin{figure}[t]
    \centering

    \begin{subfigure}[t]{0.48\linewidth}
        \centering
        \includegraphics[width=\linewidth]{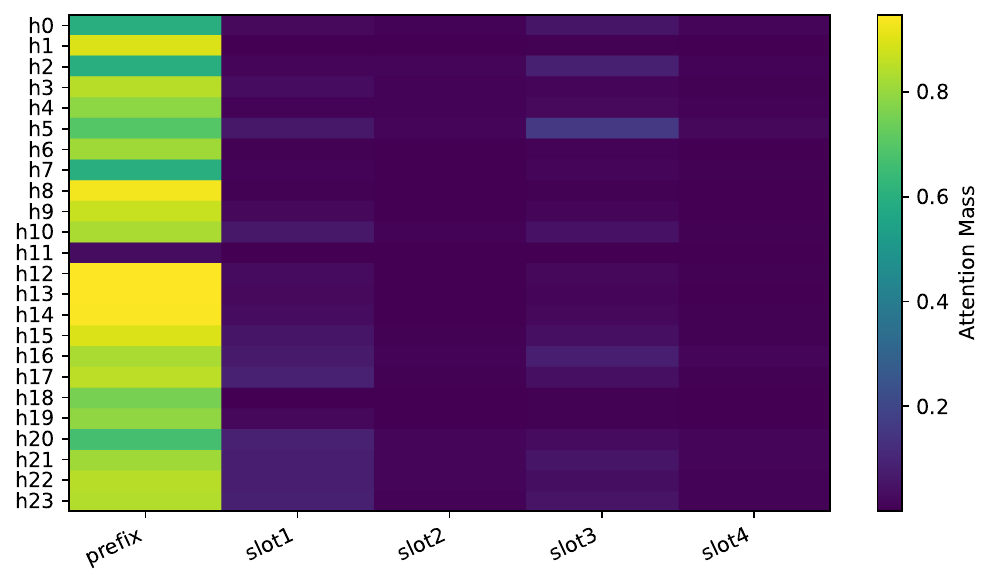}
        \caption{Dense Prefill}
        \label{fig:ap1}
    \end{subfigure}
    \hfill
    \begin{subfigure}[t]{0.48\linewidth}
        \centering
        \includegraphics[width=\linewidth]{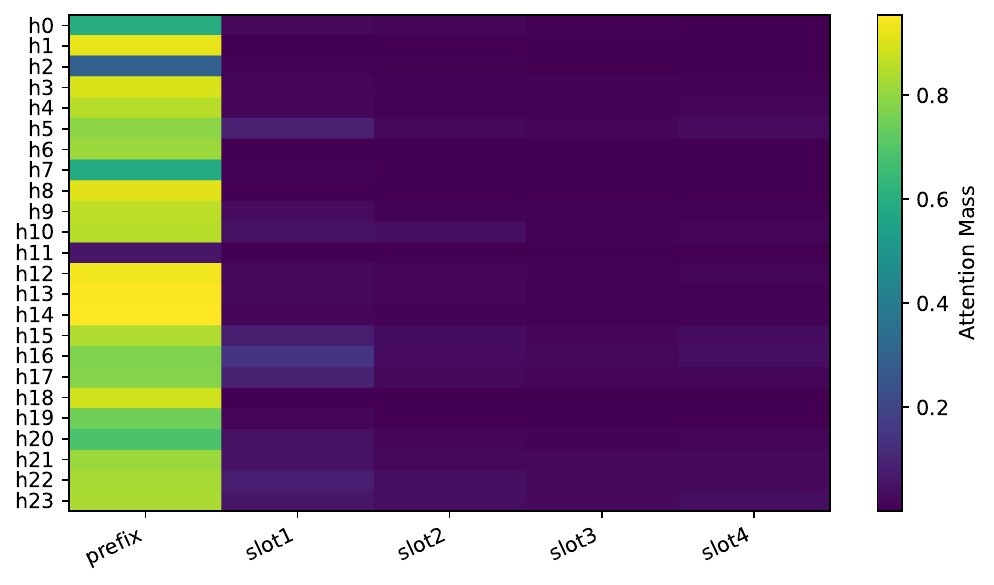}
        \caption{Na\"ive Reuse}
        \label{fig:ap2}
    \end{subfigure}

    \vspace{0.5em}

    \begin{subfigure}[t]{0.48\linewidth}
        \centering
        \includegraphics[width=\linewidth]{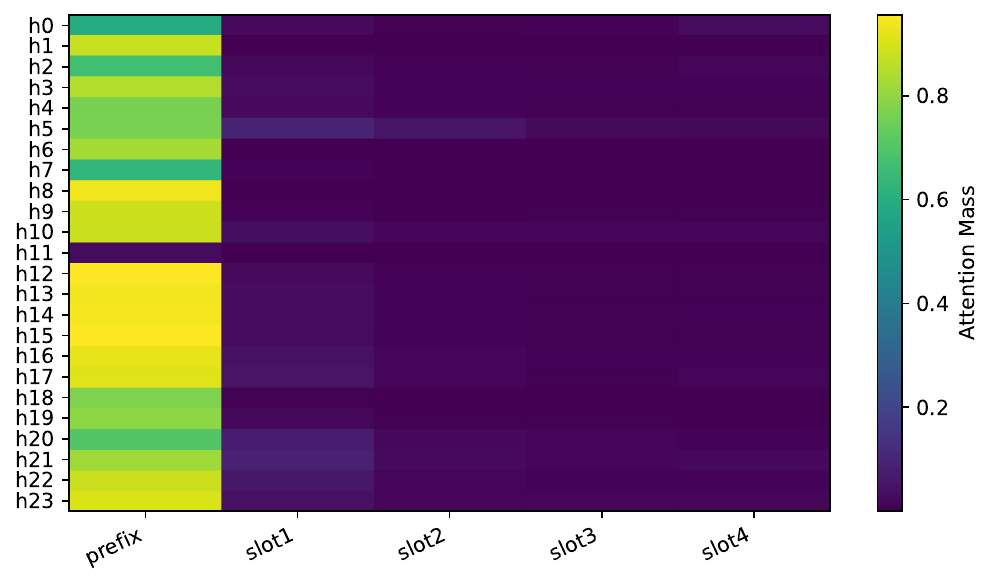}
        \caption{KVCOMM}
        \label{fig:ap3}
    \end{subfigure}
    \hfill
    \begin{subfigure}[t]{0.48\linewidth}
        \centering
        \includegraphics[width=\linewidth]{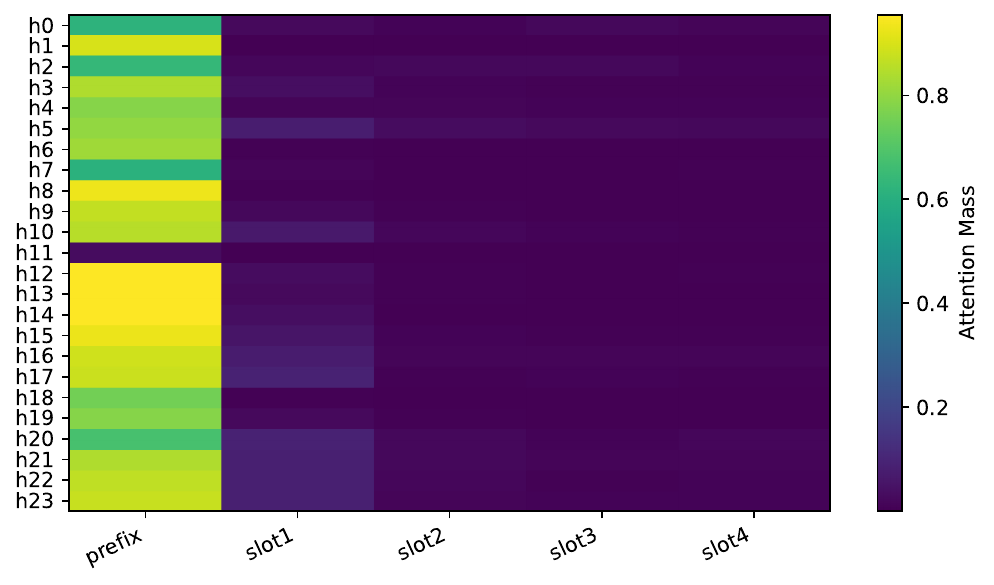}
        \caption{PAL-KV}
        \label{fig:ap4}
    \end{subfigure}
    \caption{Attention mass visualization under different KV reuse strategies.}
    \label{fig:attn_compare}
\end{figure}

\subsection{Additional Attention Visualizations}
\label{attention}
\noindent
This section provides additional attention visualizations for representative examples under dense prefill and reuse-based methods, complementing the aggregate statistics reported in the main text.
Figure~\ref{fig:attn_compare} shows token-level attention mass over the shared prefix and candidate slots during the judge's first-token generation.
While the main text summarizes the \emph{average} attention allocated to each region, here we present qualitative examples that illustrate how the attention patterns differ across methods.

In dense prefill, the judge typically allocates non-trivial attention to multiple candidate blocks (e.g., \texttt{slot1} and \texttt{slot3} in the shown example), which is consistent with the intended behavior of joint multi-candidate comparison.
In contrast, Na\"ive Reuse often exhibits a highly concentrated pattern where attention collapses onto a single early slot (frequently \texttt{slot1}), suggesting that position-only stitching can distort cross-candidate interactions and effectively reduce the judge's ability to incorporate later evidence.
KVCOMM and PAL-KV partially mitigate this collapse by restoring some attention to non-first slots; however, their attention distributions remain noticeably more peaked than dense prefill, indicating that the judge still under-attends to certain candidate blocks relative to the dense reference.
Overall, these visualizations support the interpretation that KV reuse can bias the judge toward a more \emph{shortcut} decision process---selecting a ``best'' candidate with less thorough cross-candidate comparison---which helps explain the observed drops in JCR.
% TODO: Insert per-example attention maps / bar charts and brief interpretation notes.

\section{Case Study}
\label{app:case}

Figures~\ref{fig:case1} and~\ref{fig:case2} provide a GSM8K case study under \texttt{Parallel Exploration} with \texttt{shuffle} ordering, illustrating how judge selections can change relative to dense prefill.

This case study illustrates \emph{decision non-invariance} rather than answer degradation.
Among the four candidates, Agents~1 and~4 provide essentially equivalent and correct solutions (both conclude \$12 with near-identical reasoning steps), while Agents~2 and~3 make the same arithmetic mistake (\$16).
Under dense prefill, the judge selects Agent~1, whereas under KVCOMM it selects Agent~4, despite producing the same final answer.

\textbf{Why this is still informative.}
Because multiple candidates are correct and highly similar, the judge’s choice is driven by fine-grained cross-candidate comparison and implicit tie-breaking (e.g., how evidence from each block is integrated, which rationale is attended to, and how the judge maps candidate identities under a shuffled layout), rather than by correctness alone.
This is precisely the regime where we observe large drops in JCR: KV reuse can change \emph{which} candidate the judge attributes as best even when task accuracy remains unchanged.
In other words, the failure mode here is not ``worse answers,'' but \emph{unstable attribution/selection} relative to dense prefill, which undermines auditability and accountability in judge-centric pipelines.

\section{Reproducibility Details}
\label{app:repro}

\subsection{Prompting Templates and Output Formats}
\label{app:prompts}
\noindent
This appendix summarizes the prompting templates used for both execution agents and the judge, as well as the structured output format required for reliable parsing of the final answer and the selected candidate index.

\noindent\textbf{Execution-agent prompting.}
To ensure that candidate responses are directly comparable (i.e., differences arise from stochasticity or contextual variation rather than mismatched roles), we use \emph{the same role prompt and instruction template} for all execution agents within each benchmark.
Specifically, for \textbf{MMLU} we design a dedicated role prompt (\texttt{MMLU Solver}) tailored to multiple-choice reasoning in Figure~\ref{fig:mmlu};
for \textbf{GSM8K} we reuse KVCOMM's \texttt{Math Solver} role prompt in Figure~\ref{fig:math};
and for \textbf{HumanEval} we reuse the \texttt{Programming Expert} role prompt in Figure~\ref{fig:code}.
All execution agents follow the same output schema and produce (i) a final answer and (ii) an explicit rationale, which helps the judge compare candidates under a consistent format.

\noindent\textbf{Judge prompting.}
For the judge agent, we adapt the base role prompt \texttt{FinalInfer} to a selection-oriented variant \texttt{FinalSelectBest} in Figure~\ref{fig:judge_mmlu}, ~\ref{fig:judge_math} and ~\ref{fig:judge_code}.
The judge is instructed to \emph{jointly} (a) select the best candidate by outputting its index and (b) produce the final answer, while keeping the selection decision, final answer, and justification in \emph{separate structured fields}.
This design aligns with our JCR evaluation, which focuses on whether KV reuse changes the judge's selected candidate relative to dense prefill.
\subsection{Inference Setup}
\label{app:impl}
All experiments are run on two NVIDIA RTX 4090 GPUs with 24GB memory each.
We set the maximum generation length to 512 tokens for both execution agents and the judge. All models are executed in \texttt{bfloat16}; empirically, \texttt{float16} can lead to \texttt{NaN} failures on Qwen-2.5 models in our setup.
% TODO: Add decoding configs (temperature, top-p), batching, and random seeds.

\subsection{Implementation Details of KVCOMM and PAL-KV}
\label{app:kvcomm_details}
\noindent
We provide additional implementation details for KVCOMM and PAL-KV beyond the main text, including anchor construction, matching views, correction retrieval, and fallback criteria.
\paragraph{Our main-paper description is simplified.}
In the main paper, we describe KVCOMM-style reuse as \emph{retrieving the nearest anchor and applying its cache offset}.
This is a simplified view used for exposition. In the original KVCOMM design, reuse is \emph{gated} by a shareability
criterion (a thresholded decision), and the offset is typically predicted by \emph{interpolating multiple anchors} rather
than always using a single nearest one.

\paragraph{Reuse gating (thresholded decision).}
KVCOMM maintains an anchor pool for each placeholder segment. For an incoming placeholder $\phi$, KVCOMM first decides
whether it is \emph{shareable} by checking (i) length compatibility and (ii) embedding-based proximity to existing anchors.
Concretely, the anchor prediction uses a threshold $\gamma$ to determine whether the matched anchors are sufficiently
concentrated in embedding space; if the criterion is not satisfied, KVCOMM falls back to dense prefilling and treats the
new sample as a new anchor to expand future coverage. This “reuse-or-fallback” logic is part of KVCOMM’s standard online
procedure. 
\paragraph{Offset approximation (weighted aggregation).}
When $\phi$ is predicted shareable, KVCOMM retrieves a set of matched anchors and estimates the KV offset by a
\emph{weighted sum} of their stored deviations. In particular, the weights are computed by a softmax over negative
embedding distances, so closer anchors contribute more. The approximated placeholder cache is obtained by adding the
weighted offset to the base cache; neighboring prefix segments are updated analogously.

\paragraph{No matched anchors $\Rightarrow$ no reuse.}
If no anchor satisfies the shareability criterion (e.g., no length-compatible / sufficiently close anchors), KVCOMM does
\emph{not} reuse KV for that placeholder. Instead it performs dense prefilling, measures the true deviation to the base
cache, and stores it as a new anchor entry for future requests.

\paragraph{PAL-KV: a probe that keeps KVCOMM’s reuse rate unchanged.}
Our probe method PAL-KV is designed to analyze the effect of \emph{pooling/aggregation} while keeping the \emph{reuse rate}
identical to KVCOMM. Therefore, PAL-KV \emph{never changes} KVCOMM’s reuse gating decision: we only activate PAL-KV’s
``anchor-pooling'' behavior \emph{when KVCOMM already decides the placeholder is shareable (i.e., would reuse)}.
In those reuse cases, instead of applying only a single nearest-anchor offset, PAL-KV explicitly pools a larger set of
matched anchors (the same matched set used by KVCOMM’s interpolation, or an expanded top-$K$ subset) and aggregates their
offsets via the same distance-based weighting. In contrast, when KVCOMM falls back to dense prefilling, PAL-KV also falls
back, so the frequency of reuse events is preserved by construction.

% TODO: Describe anchor pool construction, the matching view v_i, RetrieveOffset/RetrieveDelta, and reliability gating.
\subsection{Shuffle Setting and Motivation}
A potential concern is that the \texttt{shuffle} setting may appear artificial.
In practice, however, non-fixed candidate ordering is common in judge-centric multi-agent pipelines.
First, many MAS deployments use dynamic and non-canonical communication graphs (e.g., varying groupings, asynchronous
message passing, or early-stopping/anytime behaviors), which induces variability in when each candidate becomes available.
To reduce end-to-end latency, systems often stream candidates to the judge as soon as they are produced, rather than
waiting to enforce a fixed global order; consequently, the presentation order of candidate blocks can vary across runs.

Second, the judge literature has repeatedly highlighted position-related effects (e.g., position preference / order bias),
and it is therefore standard practice to evaluate robustness under alternative candidate permutations.
Our \texttt{shuffle} regime serves as a controlled and reproducible proxy for these naturally occurring order variations
and robustness checks, enabling us to isolate how KV reuse behaves under layout instability.

\subsection{Datasets Details}
We follow KVCOMM's dataset processing and evaluation protocol as closely as possible; for full preprocessing and prompting
details, please refer to our released code. We evaluate on: (i) a fixed MMLU\footnote{\url{https://huggingface.co/datasets/cais/mmlu}} validation subset of 153 questions sampled
with \texttt{seed=888}; (ii) the full GSM8K\footnote{\url{https://huggingface.co/datasets/openai/gsm8k}} test set (1,319 problems); and (iii) the full HumanEval\footnote{\url{https://huggingface.co/datasets/openai/openai_humaneval}} Python set (161 tasks).

\begin{figure*}[t]
\centering
\caption{Math Solver Prompting Template.}
\label{fig:math}
\begin{tcolorbox}
\textbf{Math Solver}\\
You are a math expert. \\
You will be given a math problem and hints from other agents. \\
Give your own solving process step by step based on hints. \\
The last line of your output contains only the final result without any units, for example: The answer is 140\\
You will be given some examples you may refer to.\\

Q: Angelo and Melanie want to plan how many hours over the next week they should study together for their test next week. \\
They have 2 chapters of their textbook to study and 4 worksheets to memorize. \\
They figure out that they should dedicate 3 hours to each chapter of their textbook and 1.5 hours for each worksheet. \\
If they plan to study no more than 4 hours each day, how many days should they plan to study total over the next week if they take a 10-minute break every hour, 
include 3 10-minute snack breaks each day, and 30 minutes for lunch each day?.\\

A: Let's think step by step. \\
Angelo and Melanie think they should dedicate 3 hours to each of the 2 chapters, 3 hours x 2 chapters = 6 hours total.\\
For the worksheets they plan to dedicate 1.5 hours for each worksheet, 1.5 hours x 4 worksheets = 6 hours total.
Angelo and Melanie need to start with planning 12 hours to study, at 4 hours a day, 12 / 4 = 3 days.\\
However, they need to include time for breaks and lunch. Every hour they want to include a 10-minute break, 
so 12 total hours x 10 minutes = 120 extra minutes for breaks.\\
They also want to include 3 10-minute snack breaks, 3 x 10 minutes = 30 minutes.\\
And they want to include 30 minutes for lunch each day, so 120 minutes for breaks + 30 minutes for snack breaks + 30 minutes for lunch = 180 minutes, or 180 / 60 minutes per hour = 3 extra hours.\\
So Angelo and Melanie want to plan 12 hours to study + 3 hours of breaks = 15 hours total.\\
They want to study no more than 4 hours each day, 15 hours / 4 hours each day = 3.75\\
They will need to plan to study 4 days to allow for all the time they need.\\
The answer is 4\\

\texttt{\{QA\_case\_2\_content\}}\\

\texttt{\{QA\_case\_3\_content\}}\\

Q: \texttt{\{user\_question\}}\\
At the same time, the output of other agents is as follows:\\ 
Agent 1, role is Math Solver, output is:\\
\texttt{\{agent\_1\_current\}}\\
Agent 2, role is Math Solver, output is:\\
\texttt{\{agent\_2\_current\}}\\
Agent 3, role is Math Solver, output is:\\
\texttt{\{agent\_3\_current\}}
\end{tcolorbox}
\end{figure*}

\begin{figure*}[t]
\centering
\caption{Programming Expert Prompting Template.}
\label{fig:code}
\begin{tcolorbox}
\textbf{Programming Expert}\\
You are a programming expert.
You will be given a function signature and its docstring by the user.
You may be able to get the output results of other agents. They may have passed internal tests, but they may not be completely correct.
Write your full implementation (restate the function signature).
Use a Python code block to write your response. For example:
\begin{verbatim}
```python
print('Hello world!')
```
\end{verbatim}
Do not include anything other than Python code blocks in your response.
Do not change function names and input variable types in tasks.\\
The task is: \texttt{\{user\_question\}}\\
At the same time, the outputs and feedbacks of other agents are as follows:\\
Agent 1 as a Programming Expert: \\
The code written by the agent is:\\
\texttt{\{agent\_1\_current\}}\\
Agent 2 as a Programming Expert: \\
The code written by the agent is:\\
\texttt{\{agent\_2\_current\}}\\
Agent 3 as a Programming Expert: \\
The code written by the agent is:\\
\texttt{\{agent\_3\_current\}}
% Whether it passes internal testing?\\
% \texttt{\{condition\_1\_current\}}\\
% Agent 2 as a Algorithm Designer provides the following info:\\
% \texttt{\{agent\_2\_current\}}\\
% In the last round of dialogue, the outputs and feedbacks of some agents were:\\
% Agent 1 as a Project Manager:\\
% The code written by the agent was:\\
% \texttt{\{agent\_1\_history\_-1\}}\\
% Whether it passed internal testing?\\
% \texttt{\{condition\_1\_history\_-1\}}\\
% Agent 2 as a Algorithm Designer provided the following info:\\
% \texttt{\{agent\_2\_history\_-1\}}
\end{tcolorbox}
\end{figure*}

\begin{figure*}[t]
\centering
\caption{MMLU Solver Prompting Template.}
\label{fig:mmlu}
\begin{tcolorbox}
% \small
\textbf{MMLU Solver} \\
Solve the MMLU multiple-choice question (A/B/C/D; only one is correct).\\
Output:\\
1) First line: ONLY one letter (A, B, C, or D).\\
2) Then a short justification (total < 100 words).\\
You may use other agents' answers as references if provided, but verify independently.\\
The Question is:\\
\texttt{\{user\_question\}}\\
At the same time, the output of other agents is as follows:\\ 
Agent 1 as a MMLU Solver: \\
\texttt{\{agent\_1\_current\}}\\
Agent 2 as a MMLU Solver: \\
\texttt{\{agent\_2\_current\}}\\
Agent 3 as a MMLU Solver: \\
\texttt{\{agent\_3\_current\}}
\end{tcolorbox}
\end{figure*}

\begin{figure*}[t]
\centering
\caption{FinalSelectBest for Math Prompting Template.}
\label{fig:judge_math}
\begin{tcolorbox}
% \small
\textbf{FinalSelectBest for Math} \\
You are the top decision-maker and judge.\\
You will be given a math problem and multiple candidate answers from different agents.\\
Your job is to select which single agent produced the best answer.\\\\
Selection criteria:\\
- Consider BOTH final answer correctness AND reasoning quality.\\
- If an answer happens to be correct but the reasoning is flawed/irrelevant, prefer the one with correct reasoning.\\
- You MUST select exactly one agent id from the candidate list.\\
Output format requirements (MUST follow exactly):\\
1) First line: Selected agent id: <id>  (id must be one of: {allowed})\\
2) Then write a short justification (any format is ok).\\
3) Last line: The answer is <number>\\
The task is: \texttt{\{user\_question\}}\\
Candidate Agent Answers:\\ 
Agent 1, role is Math Solver, output is:\\
\texttt{\{agent\_1\_current\}}\\
Agent 2, role is Math Solver, output is:\\
\texttt{\{agent\_2\_current\}}\\
Agent 3, role is Math Solver, output is:\\
\texttt{\{agent\_3\_current\}}\\
Agent 4, role is Math Solver, output is:\\
\texttt{\{agent\_4\_current\}}
\end{tcolorbox}
\end{figure*}

\begin{figure*}[t]
\centering
\caption{FinalSelectBest for MMLU Prompting Template.}
\label{fig:judge_mmlu}
\begin{tcolorbox}
% \small
\textbf{FinalSelectBest for MMLU} \\
You are a judge that selects which agent produced the best multiple-choice answer.\\
You will be given a question and candidate answers from different agents.\\
Your job is to choose the best agent and output its id and final choice.\\\\
Selection criteria:\\
- Prefer the agent with the most correct, reliable reasoning.\\
- If multiple agents choose the same option, prefer the one with better justification.\\
Output format requirements (MUST follow exactly):\\
1) First line: Selected agent id: <id> (id must be one of: {allowed})\\
2) Then write a short justification (any format is ok).\\
3) Final line: <choice> (must be exactly one of: A, B, C, D)\\
3) Do NOT output anything else.\\
The task is: \texttt{\{user\_question\}}\\
Candidate Agent Answers:\\ 
Agent 1, role is MMLU Solver, output is:\\
\texttt{\{agent\_1\_current\}}\\
Agent 2, role is MMLU Solver, output is:\\
\texttt{\{agent\_2\_current\}}\\
Agent 3, role is MMLU Solver, output is:\\
\texttt{\{agent\_3\_current\}}\\
Agent 4, role is MMLU Solver, output is:\\
\texttt{\{agent\_4\_current\}}
\end{tcolorbox}
\end{figure*}

\begin{figure*}[t]
\centering
\caption{FinalSelectBest for Code Prompting Template.}
\label{fig:judge_code}
\begin{tcolorbox}
% \small
\textbf{FinalSelectBest for Code} \\
You are a judge that selects the best agent answer.\\
Evaluation Instructions:\\
1. Examine the question closely to understand its requirements.\\
2. Read each candidate answer thoroughly and assess its relevance and accuracy about the question.\\
3. Choose the answer that most accurately and completely addresses the question.\\
4. Ignore the candidate answers if they do not give a direct answer, for example, using 'unable to ...', 'as an AI ...'.\\
5. Copy the chosen answer exactly as it is presented, maintaining its original format.\\
6. Adhere to the constraints: \\
Output format:\\
- First line: Selected agent id: <id> (choose <id> from: {joined})\\
- Then output exactly one Python code block copied from the chosen agent.\\
- Do not add any explanation outside the code block.\\
Note: If none of the answers fully meet the question's criteria, select the one closest to fulfilling them.\\
The task is: \texttt{\{user\_question\}}\\
Candidate Answers for Evaluation:\\ 
Agent 1, role is Programming Expert, output is:\\
\texttt{\{agent\_1\_current\}}\\
Agent 2, role is Programming Expert, output is:\\
\texttt{\{agent\_2\_current\}}\\
Agent 3, role is Programming Expert, output is:\\
\texttt{\{agent\_3\_current\}}\\
Agent 4, role is Programming Expert, output is:\\
\texttt{\{agent\_4\_current\}}
\end{tcolorbox}
\end{figure*}

\begin{figure*}[t]
\centering
\caption{Case Study: A GSM8K Example under Parallel Exploration, Shuffle Settings, Part 1 Candidate outputs.}
\label{fig:case1}
\begin{tcolorbox}
\textbf{Exemples} \\
Question:\\
John invited 20 people to a birthday party.  Each guest will eat 2 hot dogs.  He already has 4 hot dogs left over from a previous party.  If a pack of hot dogs contains 6 hot dogs and costs \$2, how much does he need to spend on hot dogs?\\
Candidate Agent Answers:\\
Agent 1:\\
Let's think step by step\\
John invited 20 people to his birthday party, and each guest will eat 2 hot dogs. 
So, the total number of hot dogs needed is 20 people * 2 hot dogs/person = 40 hot dogs.\\
John already has 4 hot dogs left over from a previous party, so he needs 40 hot dogs - 4 hot dogs = 36 hot dogs.\\
Since a pack of hot dogs contains 6 hot dogs, John needs 36 hot dogs / 6 hot dogs/pack = 6 packs of hot dogs.\\
Since each pack of hot dogs costs \$2, John needs to spend 6 packs * \$2/pack = \$12 on hot dogs.\\
The answer is 12\\
Agent 3:\\
Let's think step by step.\\
First, we need to determine how many hot dogs John will need for the party. \\
He invited 20 people, and each guest will eat 2 hot dogs, so 20 x 2 = 40 hot dogs.\\
He already has 4 hot dogs left over, so he needs 40 + 4 = 44 hot dogs.\\
Since a pack of hot dogs contains 6 hot dogs, we need to divide 44 by 6 to find out how many packs John needs to buy. \\
44 / 6 = 7.33, so John will need to buy 8 packs of hot dogs (since you can't buy a fraction of a pack).\\
Each pack costs \$2, so 8 packs will cost 8 x 2 = \$16.
The answer is 16\\
Agent 2:\\
Let's think step by step.\\
First, we need to determine how many hot dogs John will need for the party.\\ 
He invited 20 people, and each guest will eat 2 hot dogs, so 20 x 2 = 40 hot dogs.\\
John already has 4 hot dogs left over, so he needs 40 + 4 = 44 hot dogs.\\
Since a pack of hot dogs contains 6 hot dogs, we need to divide 44 by 6 to find out how many packs John needs to buy.\\
44 / 6 = 7.33, so John will need to buy 8 packs of hot dogs (since you can't buy a fraction of a pack).\\
Each pack costs \$2, so 8 x \$2 = \$16.\\
The answer is 16\\
Agent 4:\\
Let's think step by step.\\
First, we need to determine the total number of hot dogs needed for the party. \\
There are 20 guests, and each guest will eat 2 hot dogs, so 20 x 2 = 40 hot dogs are needed.\\
John already has 4 hot dogs left over, so he needs 40 - 4 = 36 hot dogs.\\
Since a pack of hot dogs contains 6 hot dogs, John will need to buy 36 / 6 = 6 packs of hot dogs.\\
Each pack of hot dogs costs \$2, so John will need to spend 6 x 2 = \$12 on hot dogs.\\
The answer is 12\\
\end{tcolorbox}
\end{figure*}

\begin{figure*}[t]
\centering
\caption{Case Study: A GSM8K Example under Parallel Exploration, Shuffle Settings, Part2 Judge outputs.}
\label{fig:case2}
\begin{tcolorbox}
\textbf{Exemples} \\
Question:\\
John invited 20 people to a birthday party.  Each guest will eat 2 hot dogs.  He already has 4 hot dogs left over from a previous party.  If a pack of hot dogs contains 6 hot dogs and costs \$2, how much does he need to spend on hot dogs?\\

Dense Prefill Output:\\
Selected agent id: 1\\
The selected agent provided a clear and step-by-step reasoning for the problem. Both the final answer and the reasoning are correct, but the reasoning is more concise and directly addresses the problem. The agent correctly calculates the number of hot dogs needed, subtracts the leftover hot dogs, and then divides the result by the number of hot dogs per pack to find the number of packs needed. The final answer is also correct, and the agent correctly multiplies the number of packs by the cost per pack to find the total cost.\\
The answer is 12\\

KVCOMM Output:\\
Selected agent id: 4\\
The selected agent provided a clear and step-by-step reasoning for the problem. The solution correctly computes the total hot dogs needed (20 guests × 2 each), accounts for the 4 hot dogs already available by subtracting them, and then converts the remaining requirement into packs by dividing by 6 hot dogs per pack. Finally, it multiplies the number of packs by the \$2 cost per pack to get the total spending. The reasoning is well-structured, easy to follow, and the final answer is correct.\\
The answer is 12\\
\end{tcolorbox}
\end{figure*}
\end{document}